\documentclass[ALICE,manyauthors]{cernphprep}
\usepackage{hyperref}
\usepackage{lineno}
\usepackage{xcolor}
\usepackage{euscript}
\usepackage{cite}
\usepackage[scientific-notation=true]{siunitx}
\usepackage[greek,english]{babel}
\usepackage{multirow}

\usepackage[Symbolsmallscale]{upgreek}
\usepackage{multirow}

\begin{document}
%

\newcommand{\pp}           {pp\xspace}
\newcommand{\ppbar}        {\mbox{$\mathrm {p--\overline{p}}$}\xspace}
\newcommand{\XeXe}         {\mbox{Xe--Xe}\xspace}
\newcommand{\PbPb}         {\mbox{Pb--Pb}\xspace}
\newcommand{\pAa}           {\mbox{pA}\xspace}
\newcommand{\pPb}          {\mbox{p--Pb}\xspace}
\newcommand{\AuAu}         {\mbox{Au--Au}\xspace}
\newcommand{\dAu}          {\mbox{d--Au}\xspace}

\newcommand{\snn}          {\ensuremath{\sqrt{s_{\mathrm{NN}}}}\xspace}
\newcommand{\pt}           {\ensuremath{p_{\rm T}}\xspace}
\newcommand{\meanpt}       {$\langle p_{\mathrm{T}}\rangle$\xspace}
\newcommand{\ycms}         {\ensuremath{y_{\rm CMS}}\xspace}
\newcommand{\ylab}         {\ensuremath{y_{\rm lab}}\xspace}
\newcommand{\etarange}[1]  {\mbox{$\left | \eta \right |~<~#1$}}
\newcommand{\yrange}[1]    {\mbox{$\left | y \right |~<~#1$}}
\newcommand{\dndy}         {\ensuremath{\mathrm{d}N_\mathrm{ch}/\mathrm{d}y}\xspace}
\newcommand{\dndeta}       {\ensuremath{\mathrm{d}N_\mathrm{ch}/\mathrm{d}\eta}\xspace}
\newcommand{\avdndeta}     {\ensuremath{\langle\dndeta\rangle}\xspace}
\newcommand{\dNdy}         {\ensuremath{\mathrm{d}N_\mathrm{ch}/\mathrm{d}y}\xspace}
\newcommand{\Npart}        {\ensuremath{N_\mathrm{part}}\xspace}
\newcommand{\Ncoll}        {\ensuremath{N_\mathrm{coll}}\xspace}
\newcommand{\dEdx}         {\ensuremath{\textrm{d}E/\textrm{d}x}\xspace}
\newcommand{\RpPb}         {\ensuremath{R_{\rm pPb}}\xspace}

\newcommand{\nineH}        {$\sqrt{s}~=~0.9$~Te\kern-.1emV\xspace}
\newcommand{\seven}        {$\sqrt{s}~=~7$~Te\kern-.1emV\xspace}
\newcommand{\onethree}        {$\sqrt{s}~=~13$~Te\kern-.1emV\xspace}
\newcommand{\twoH}         {$\sqrt{s}~=~0.2$~Te\kern-.1emV\xspace}
\newcommand{\twosevensix}  {$\sqrt{s}~=~2.76$~Te\kern-.1emV\xspace}
\newcommand{\five}         {$\sqrt{s}~=~5.02$~Te\kern-.1emV\xspace}
\newcommand{\twosevensixnn}{$\sqrt{s_{\mathrm{NN}}}~=~2.76$~Te\kern-.1emV\xspace}
\newcommand{\fivenn}       {$\sqrt{s_{\mathrm{NN}}}~=~5.02$~Te\kern-.1emV\xspace}
\newcommand{\LT}           {L{\'e}vy-Tsallis\xspace}
\newcommand{\GeVc}         {Ge\kern-.1emV/$c$\xspace}
\newcommand{\MeVc}         {Me\kern-.1emV/$c$\xspace}
\newcommand{\GeVmass}      {Ge\kern-.1emV/$c^2$\xspace}
\newcommand{\MeVmass}      {Me\kern-.1emV/$c^2$\xspace}
\newcommand{\lumi}         {\ensuremath{\mathcal{L}}\xspace}

\newcommand{\ITS}          {\rm{ITS}\xspace}
\newcommand{\TOF}          {\rm{TOF}\xspace}
\newcommand{\ZDC}          {\rm{ZDC}\xspace}
\newcommand{\ZDCs}         {\rm{ZDCs}\xspace}
\newcommand{\ZNA}          {\rm{ZNA}\xspace}
\newcommand{\ZNC}          {\rm{ZNC}\xspace}
\newcommand{\SPD}          {\rm{SPD}\xspace}
\newcommand{\SDD}          {\rm{SDD}\xspace}
\newcommand{\SSD}          {\rm{SSD}\xspace}
\newcommand{\TPC}          {\rm{TPC}\xspace}
\newcommand{\TRD}          {\rm{TRD}\xspace}
\newcommand{\VZERO}        {\rm{V0}\xspace}
\newcommand{\VZEROA}       {\rm{V0A}\xspace}
\newcommand{\VZEROC}       {\rm{V0C}\xspace}
\newcommand{\Vdecay} 	   {\ensuremath{V^{0}}\xspace}

\newcommand{\ee}           {\ensuremath{e^{+}e^{-}}} 
\newcommand{\pip}          {\ensuremath{\pi^{+}}\xspace}
\newcommand{\pim}          {\ensuremath{\pi^{-}}\xspace}
\newcommand{\kap}          {\ensuremath{\rm{K}^{+}}\xspace}
\newcommand{\kam}          {\ensuremath{\rm{K}^{-}}\xspace}
\newcommand{\pbar}         {\ensuremath{\rm\overline{p}}\xspace}
\newcommand{\kzero}        {\ensuremath{{\rm K}^{0}_{\rm{S}}}\xspace}
\newcommand{\lmb}          {\ensuremath{\Lambda}\xspace}
\newcommand{\almb}         {\ensuremath{\overline{\Lambda}}\xspace}
\newcommand{\prot}         {\ensuremath{\rm{p}}\xspace}
\newcommand{\aprot}         {\ensuremath{\rm{\overline{p}}}\xspace}
\newcommand{\n}         {\ensuremath{\rm{n}}\xspace}
\newcommand{\an}         {\ensuremath{\rm{\overline{n}}}\xspace}

\newcommand{\Om}           {\ensuremath{\Omega^-}\xspace}
\newcommand{\Mo}           {\ensuremath{\overline{\Omega}^+}\xspace}
\newcommand{\X}            {\ensuremath{\Xi^-}\xspace}
\newcommand{\Ix}           {\ensuremath{\overline{\Xi}^+}\xspace}
\newcommand{\Xis}          {\ensuremath{\Xi^{\pm}}\xspace}
\newcommand{\Oms}          {\ensuremath{\Omega^{\pm}}\xspace}
\newcommand{\SigZ}            {\ensuremath{\Sigma^0}\xspace}
\newcommand{\aSigZ}            {\ensuremath{\overline{\Sigma^0}}\xspace}
\newcommand{\antik}   {$\mathrm{\overline{K}}\,$}

\newcommand{\Ledn}         {Lednick\'y--Lyuboshits\xspace}
\newcommand{\chiEFT}       {\ensuremath{\chi}\rm{EFT}\xspace}
\newcommand{\ks}     {\ensuremath{k^{*}}\xspace}
\newcommand{\rs}     {\ensuremath{r^{*}}\xspace}
\newcommand{\mt}     {\ensuremath{m_{\mathrm{T}}}\xspace}
\newcommand{\Cth}           {C_\mathrm{th}\xspace}
\newcommand{\Cexp}           {C_\mathrm{exp}\xspace}
\newcommand{\CF}           {\ensuremath{C(\ks)}\xspace}
\newcommand{\Sr}            {\ensuremath{S(\rs)}\xspace}
\newcommand{\BBar}            {\ensuremath{\rm{B}\mbox{--}\rm{\overline{B}}}\xspace}
\newcommand{\pap}            {\ensuremath{\prot\mbox{--}\aprot}\xspace}
\newcommand{\paL}            {\ensuremath{\prot\mbox{--}\almb}\xspace}
\newcommand{\apL}            {\ensuremath{\aprot\mbox{--}\lmb}\xspace}
\newcommand{\LaL}            {\ensuremath{\lmb\mbox{--}\almb}\xspace}
\newcommand{\Km}            {\ensuremath{\rm K^-\mbox{--}\prot}\xspace}
\newcommand{\patp}            {\ensuremath{\prot\mbox{--}\aprot}\xspace}
\newcommand{\LatL}            {\ensuremath{\lmb\mbox{--}\almb}\xspace}
\newcommand{\patL}            {\ensuremath{\prot\mbox{--}\almb}\xspace}
\newcommand{\pprot}            {\ensuremath{\prot\mbox{--}\prot}\xspace}
\newcommand{\nan}            {\ensuremath{\n\mbox{--}\an}\xspace}

\newcommand{\NNbar}            {\ensuremath{\mathrm{N\mbox{--}{\overline{N}}}}\xspace}
\newcommand{\NYbar}            {\ensuremath{\mathrm{N\mbox{--}{\overline{Y}}}}\xspace}
\newcommand{\YYbar}            {\ensuremath{\mathrm{Y\mbox{--}{\overline{Y}}}}\xspace}
\newcommand{\Swzero}        {\ensuremath{\rm^1S_\mathrm{0}}\xspace}
\newcommand{\Swtre}        {\ensuremath{\rm^3S_\mathrm{1}}\xspace}
\newcommand{\Pwtrezero}        {\ensuremath{\rm^3P_\mathrm{0}}\xspace}
\newcommand{\Pwunouno}        {\ensuremath{\rm^1P_\mathrm{1}}\xspace}
\newcommand{\Pwtreuno}        {\ensuremath{\rm^3P_\mathrm{1}}\xspace}
\newcommand{\Pwtredue}        {\ensuremath{\rm^3P_\mathrm{2}}\xspace}

\newcommand{\Imscatt}            {\ensuremath{\mathcal{I}f_0}\xspace}
\newcommand{\Rescatt}            {\ensuremath{\mathcal{R}f_0}\xspace}
\newcommand{\effran}            {\ensuremath{d_0}\xspace}

\newcommand{\SwzeroResult} {\ensuremath{\omega_{\Swzero} = 1.19 \pm 0.10\, \mathrm{(stat)} \,{\pm 0.19} \, \mathrm{(syst)}}\xspace}
\newcommand{\PwtrezeroResult} {\ensuremath{\omega_{\Pwtrezero} = 40.04  \pm 4.06\, \mathrm{(stat)} \,{\pm 4.24} \, \mathrm{(syst)}}\xspace}
\newcommand{\ImscattpaL} {\ensuremath{2.59 ^{\pm 0.13\, \mathrm{(stat)}} _{{\pm 0.33} \, \mathrm{(syst)}}}\xspace}
\newcommand{\effranpaL} {\ensuremath{3.39 ^{\pm 0.27\, \mathrm{(stat)}}\xspace}}
\newcommand{\ImscattLaL} {\ensuremath{0.72 ^{\pm 0.16\, \mathrm{(stat)}} _{{\pm 0.23} \, \mathrm{(syst)}}}\xspace}
\newcommand{\effranLaL} {\ensuremath{2.96 ^{\pm 0.30\, \mathrm{(stat)}} _{{\pm 0.51} \, \mathrm{(syst)}}}\xspace}

\newcommand{\ImscattpaLPb} {\ensuremath{0.53 ^{\pm 0.04\, \mathrm{(stat)}} _{{\pm 0.15} \, \mathrm{(syst)}}}\xspace}
\newcommand{\effranpaLPb} {\ensuremath{3.06 ^{\pm 0.14\, \mathrm{(stat)}} _{{\pm 0.98} \, \mathrm{(syst)}}}\xspace}
\newcommand{\ImscattLaLPb} {\ensuremath{0.40 ^{\pm 0.06\, \mathrm{(stat)}} _{{\pm 0.18} \, \mathrm{(syst)}}}\xspace}
\newcommand{\effranLaLPb} {\ensuremath{2.76 ^{\pm 0.29\, \mathrm{(stat)}} _{{\pm 0.73} \, \mathrm{(syst)}}}\xspace}

\begin{titlepage}
\PHyear{2021}       
\PHnumber{083}      
\PHdate{10 May}  
\title{Investigating the role of strangeness in baryon--antibaryon annihilation at the LHC}
\ShortTitle{Investigating the role of strangeness in baryon--antibaryon annihilation}   

\Collaboration{ALICE Collaboration\thanks{See Appendix~\ref{app:collab} for the list of collaboration members}}
\ShortAuthor{ALICE Collaboration} 

\begin{abstract}
Annihilation dynamics plays a fundamental role in the baryon--antibaryon interaction (\BBar) at low-energy and its strength and range are crucial in the assessment of possible baryonic bound states. 
Experimental data on annihilation cross sections are available for the \patp system but not in the low relative momentum region.
Data regarding the \BBar interaction with strange degrees of freedom are extremely scarce, hence the modeling of the annihilation contributions is mainly based on nucleon--antinucleon (\NNbar) results, when available.
In this letter we present a measurement of the \patp, $\paL\oplus\apL$ and \LatL interaction using correlation functions in the relative momentum space in high-multiplicity triggered pp collisions at \onethree recorded by ALICE at the LHC.
In the \pap system  the couplings to the mesonic channels in different partial waves are extracted by adopting a coupled-channel approach with recent \chiEFT potentials. The inclusion of these inelastic channels provides good agreement with the data, showing a significant presence of the annihilation term down to zero momentum. 
Predictions obtained using the \Ledn formula and scattering parameters obtained from
heavy-ion collisions, hence mainly sensitive to elastic processes, are compared with the experimental $\paL\oplus\apL$ and \LaL correlations. The model describes the \LaL data and underestimates the $\paL\oplus\apL$ data in the region of momenta below 200 \MeVc.
The observed deviation indicates a different contribution of annihilation channels to the two systems containing strange hadrons.
\end{abstract}
\end{titlepage}

\setcounter{page}{2} 

\section{Introduction and physics motivation}
\noindent
The baryon--antibaryon interaction (\BBar) is dominated at low energies by annihilation processes, in which transitions from a state, typically composed of only mesons, to a \BBar state and vice versa are occurring.
Since the first measurement of the proton-antiproton (\pap) cross section~\cite{Chamberlain:1956nt}, a rich sample of experimental data has become available, mainly in the nucleon--antinucleon (\NNbar) sector. 
Low-energy scattering experiments~\cite{Scatt1,Scatt2,Zhou:2012ui} delivered data on the total cross section, on the elastic ($\rm{p} \overline{\rm{p}}\rightarrow \rm{p} \overline{\rm{p}}$) and charge-exchange
($\rm{p} \overline{\rm{p}}\rightarrow \rm{n} \overline{\rm{n}}$)
cross sections, down to laboratory momenta $p_{\rm lab}\approx 200$ \MeVc.
Measurements of the annihilation cross section reach even lower momenta but are affected by significant uncertainties and in particular the momentum region close to the \pap threshold is currently lacking any experimental constraint. This region is however of particular interest for the theoretical modeling of \patp interaction since the interplay between the Coulomb and the annihilation dynamics is dominant.\\
At threshold, measurements of the energy level shifts and widths of \pap atoms~\cite{KlemptProtonium} enabled the extraction of the spin-averaged scattering parameters, confirming a non-zero imaginary part of the scattering length related to the presence of inelastic channels due to the annihilation processes.\\
Great effort was made in the theoretical description of the short-range interaction (below 1 fm) of \pap systems since a stronger elastic attraction, with respect to the \pprot case, is expected to occur in some spin-isospin channels, leading to predictions of the existence of bound states (baryonia)~\cite{KlemptProtonium,Loiseaubaryonia}.
Findings of broad resonances and enhancements in the \pap invariant mass~\cite{Invmass1,Invmass2,Invmass3,Invmass4} measured in the decays of charmed and bottom mesons were reported but no clear evidence of such bound states has been found yet.
A precise understanding of the annihilation dynamics is required to assess the existence of such states since the bound spectrum could be washed out by the \BBar annihilation part of the interaction. 
The annihilation term in the \NNbar sector is typically described in chiral-effective potentials~\cite{Dai:2017ont}, meson-exchange~\cite{BBar1,BBar2,BBar5} and quark models~\cite{BBar8} by means of phenomenological optical potentials and contact interactions with parameters to be fixed from the available data.
The search for baryonia states for \BBar systems in the strangeness sector with hyperons (Y) and antihyperons ($\rm\overline{Y}$), e.g.\ \NYbar, \YYbar is even more challenging since the experimental informations are very scarce, with $\rm{p} \overline{\rm{p}} \rightarrow \Lambda \overline{\Lambda}$ being the only measured strangeness exchange process~\cite{KlemptProtonium}. 
Consequently, the modeling of the annihilation for systems as \YYbar (e.g.\ \LatL) is mainly based on the \NNbar interaction~\cite{BBar3,BBar4,BBar6,BBar7}. Measurements of the \paL invariant mass spectra in photoproduction processes $\gamma \prot \rightarrow \lmb\almb\prot$ will become available in the next years~\cite{GluexExp}, but currently no experimental informations neither theoretical predictions are present for \BBar interactions involving a nucleon (antinucleon) and an antihyperon (hyperon) such as \patL.\\
The study of annihilation in \BBar systems with strangeness is also of great interest for the modeling of the re-scattering phase in heavy-ion collisions (HIC). Several observables as particle spectra and yields strongly depend on the processes occuring at this stage of the HIC evolution, the \BBar annihilation processes above all.
Currently in HIC, the annihilation interaction for pairs containing strangeness is either modeled assuming scattering parameters similar to \pap or  with an ad-hoc suppression of the cross section with respect to the \pap counterpart~\cite{KisielBBarSTAR,Seifert:2017oyb}.\\
The present theoretical understanding of the \BBar interaction requires additional precise data particularly in the low-momentum region, where the inelastic contributions from annihilation are relevant. This would shed light on the presence of baryonia bound states and on how the annihilation dynamics changes for systems with strangeness.\\
A step in this direction has recently been  achieved with the measurements of two-particle correlations in the momentum space for \pap, $\paL$ and \LaL pairs performed in ultra-relativistic \PbPb collisions at LHC~\cite{WUTPaper}. 
The extracted spin-averaged scattering parameters are in agreement for all \BBar pairs indicating that the annihilation part for all \BBar pairs is similar at the same relative momentum. The \paL pairs were also measured in \AuAu collisions at RHIC~\cite{STARBBar}, but these results might be biased by the neglected residual correlations~\cite{KisielBBarSTAR}.
Measurements of hadron--hadron correlations have been performed in small colliding systems such as \pp and \pPb, and they delivered the most precise data on baryon--baryon and meson--baryon pairs, enabling access to the short-range strong interaction~\cite{ALICE:Run1,ALICE:LL,ALICE:pK,ALICE:pXi,ALICE:pSig0,ALICE:pOmega,ALICE:Source}. This kind of measurements in \pp collisions can probe inter-particle distances of around 1 fm and are sensitive to the presence of inelastic channels, below and above threshold~\cite{ALICE:pK,Kamiya:2019uiw,Haidenbauer:2018jvl}.\\ 
In this letter we present the measurements of the correlation functions of \pap, $\paL\oplus\apL$ and \LaL pairs in pp collisions at \onethree with the ALICE detector~\cite{ALICE,ALICEperf}. To better constrain the interaction, a differential analysis in pair-transverse-mass (\mt) intervals has been performed for the $\paL\oplus\apL$ and \LaL pairs. The work presented in this Letter delivers the most precise data at low momenta for \patp, \patL and \LatL systems and provides additional experimental constraints for the modeling of the \BBar interaction.

\section{Data analysis}\label{sec:DataAnalysis}
\noindent
The main ALICE subdetectors~\cite{ALICE,ALICEperf} used in this analysis are: the V0 detectors~\cite{Abbas:2013taa} used as trigger detectors,
the Inner Tracking System (ITS)~\cite{ALICEITS}, 
the Time Projection Chamber (TPC)~\cite{ALICETPC} and the Time-of-Flight (TOF) detector
\cite{ALICETOF}. The last three are used to track and identify charged particles. 
The high-multiplicity (HM) sample used in this analysis corresponds to 0.17\% of all inelastic pp collisions with at least one measured charged particle within $|\eta|<1$ (referred to as INEL$>$0)~\cite{ALICE:pSig0,ALICE:pOmega}. 
The corresponding HM trigger is defined by coincident hits in both \VZERO detectors synchronous with the collider bunch crossing and by requiring as well that the sum of the measured signal amplitudes in the \VZERO exceeds a multiple of the average value in minimum bias collisions. 
The rejection of pile-up events have been applied by evaluating the presence of additional event vertices as done in ~\cite{ALICE:pSig0,ALICE:Source} and a total of \num{1.0e9} HM events are selected.\\
Protons and antiprotons are reconstructed using the procedure described in Refs.~\cite{ALICE:pSig0,ALICE:Source}.
Primary protons and antiprotons are selected in the transverse momentum range $0.5 < \pt < 4.05$\,\GeVc and pseudorapidity $|\eta| < 0.8$. A minimum of 80 out of the 159 available spatial points (hits) inside the TPC are required to obtain high-quality tracks.
The TPC and TOF detectors select \prot (\aprot) candidates by the deviation $n_\sigma$ between the signal hypothesis for the considered particle and the experimental measurement, normalized by the detector resolution $\sigma$~\cite{ALICE:pSig0,ALICE:Source}.
For candidates with $p < 0.75$\,\GeVc, the particle identification (PID) is performed with the \TPC only. For larger momenta, the PID information of \TPC and \TOF are combined. The candidates are accepted if their $|n_\sigma| < 3$.
To reject non-primary protons (antiprotons), the distance of closest approach (DCA) of the candidates tracks to the primary vertex is required to be less than 0.1~cm in the $xy$-plane and less than 0.2~cm along the beam axis.
Contributions of secondary (anti)protons stemming from weak decays and misidentified candidates are extracted using Monte Carlo (MC) template fits to the measured distance of closest approach (DCA) distributions of the  to the primary vertex~\cite{ALICE:Run1}. 
The resulting \prot (\aprot) purity is $99.4\%$ ($98.9\%$). The corresponding fraction of primary particles is $82.2\%$ ($82.3\%$).\\
The reconstruction of the \lmb (\almb) candidates, via their weak decay $\lmb\rightarrow \mathrm{p}\uppi^-$ ($\almb\rightarrow \aprot\uppi^+$)~\cite{PDG}, is performed following the procedures described in Refs.~\cite{ALICE:pSig0, ALICE:Source}.
A final selection is applied based on the reconstructed invariant mass~\cite{ALICE:pSig0,ALICE:Source}.
The obtained \lmb (\almb) purity is $95.2\%$ ($96.1\%$).
Primary and secondary contributions for \lmb and \almb are extracted in a similar way as for protons,
via fits to the cosine of the pointing angle distributions using MC templates.
The fraction of primary \lmb (\almb) hyperons is about $57\%$. Secondary contributions from weak decays of neutral and charged $\Xi$ baryons amount to $22\%$. The remaining fractions are attributed to \SigZ (\aSigZ) particles.
Systematic uncertainties on the data are evaluated by varying the kinematic and topological selection criteria following~\cite{ALICE:pSig0,ALICE:Source}.

\section{Analysis of the correlation function}\label{sec:AnalysisCF}
\noindent
The main observable in the analysis presented here is the two-particle correlation function \CF, which depends on the relative momentum $\ks$ evaluated in the pair rest frame~\cite{ALICE:Run1}.
In femtoscopy measurements, the final state is fixed to the measured particle pair and the corresponding correlation function is sensitive to all the available initial, elastic and inelastic, channels produced in the collision~\cite{Haidenbauer:2018jvl,Kamiya:2019uiw}.
For the study of the \BBar interaction, the single-channel Koonin-Pratt equation~\cite{Lisa:2005dd} has to be modified in order to accommodate the inelastic contributions stemming from the annihilation channels~\cite{Haidenbauer:2018jvl,Kamiya:2019uiw}.\\
Assuming that the interaction of the pair in the final state $i$ is affected by the inelastic channels $j$, the Koonin-Pratt formula is modified by the introduction of an additive term related to the processes $j \rightarrow i$~\cite{Lednicky:1981su,Haidenbauer:2018jvl,Kamiya:2019uiw}:

\begin{align}\label{eq:corrfun_CC}
    C_i(k^*) = \int d^3 r^* S (r^*) |\psi_i (k^*,r^*)|^2 + \sum_{j\neq i} ^N \omega_j\int d^3 r^* S (r^*) |\psi_j (k^*,r^*)|^2.
\end{align}

The first integral on the right-hand side describes the elastic contribution where initial and final state coincide, while the second integral is responsible for the remaining inelastic processes $j\rightarrow i$. This last integral depends on two main ingredients: the wave function $\psi_j (k^*,r^*)$ for channel $j$ going to the final state $i$ and the conversion weights $\omega_j$. These latter quantities can be written as $\omega_j=\omega_j ^s \times \omega ^{\rm prod}_j$, in which $\omega ^{\rm prod}_j$ is related to the amount of $j$ pairs produced in the initial collision and kinematically available to be converted to the final measured state. Quantitative estimates on these production weights can be obtained combining thermal model calculations of particle yields~\cite{TFist} with kinematics constraints from transport models~\cite{Pierog:2013ria}.
If the assumed inelastic wave function $\psi_j (k^*,r^*)$ is properly accounting for the coupling strength, the corresponding $\omega_j ^s$ weight is equal to unity.\\
Recent femtoscopic measurements by the ALICE Collaboration performed on \Km pairs in \pp~\cite{ALICE:pK} and in \PbPb collisions~\cite{ALICE:pKPb} showed that by changing the colliding system, and hence the size of the emitting source $S (r^*)$, the effects on the \CF due to the inelastic contributions given by the last term in Eq.~(\ref{eq:corrfun_CC}) are enhanced or suppressed.
The wave functions $\psi_j (k^*,r^*)$ related to the inelastic channels are localized at distances \rs approximately below 1.5-2 fm and equal to zero above. Hence, performing femtoscopic measurements with a large emitting source, as it occurs in central heavy-ion collisions ($\rs$ above 5 fm), results in a correlation function mainly dominated by the elastic contribution, given by the first term of Eq.~(\ref{eq:corrfun_CC}). 
For this reason, the \CF measured in \PbPb can be modeled with the single-channel \Ledn formula~\cite{Lednicky:1981su} assuming a complex scattering length $f_0$, in which the imaginary part \Imscatt accounts for an average inelastic contribution from all $j$ channels.\\
These inelastic contributions become more relevant when performing the same measurement in small colliding systems as \pp, where the emitting source size is of the order of 1 fm~\cite{femtoreview} and the modeling of the \CF requires the knowledge of the exact elastic $\psi_i (k^*,r^*)$ and inelastic $\psi_j (k^*,r^*)$ wave functions obtained from the solution of a coupled-channel approach~\cite{femtoreview}. If the theoretical modeling of the interaction properly accounts for the inelastic channels ($\omega_j ^s = 1$), the use of the modified Koonin-Pratt formula in Eq.~(\ref{eq:corrfun_CC}) with a proper estimate of the production weights $\omega ^{\rm prod}_j$ will describe the data in both small and large colliding systems as shown in~\cite{ALICE:pKPb} for the \Km system. The use of the single-channel \Ledn model will only be applicable if the wave functions $\psi_j (k^*,r^*)$ would be strongly suppressed, corresponding to a very weak coupling to the inelastic channels.\\
The \BBar interaction investigated in this work is less known with respect to the \Km case in~\cite{ALICE:pK,ALICE:pKPb}, hence two different approaches have been used to calculate the theoretical correlation for the \pap and the \LaL, \paL pairs, respectively. For both approaches the CATS framework is used~\cite{CATS}.\\
The genuine \pap correlation is modeled either by assuming a Coulomb-only interaction or by also including a strong interaction from \NNbar chiral effective (\chiEFT) potentials at next-to-next-to-next-to-leading order ($\rm N^3$LO)~\cite{Dai:2017ont}. 
The \pap wave functions, available for S (\Swzero,\Swtre) and P (\Pwunouno,\Pwtrezero,\Pwtreuno,\Pwtredue) partial-waves (PW), have been evaluated within a coupled-channel formalism in which only the coupling to the charge-exchange \nan channel is explicitly included.
The formula in Eq.~(\ref{eq:corrfun_CC}) is used for the genuine \pap correlation function with the chiral wave functions for the elastic $i=\pap$ and the charge-exchange channel $j=\nan$~\cite{Dai:2017ont}. The wave functions $\psi^{\rm PW}_{X\rightarrow p\overline p}$, accounting for the multi-meson annihilation channels $j=X$, are not currently available. The annihilation contribution is implicitly present in the \chiEFT potentials in~\cite{Dai:2017ont} since the parameters of the model are constrained to the most-recent partial-wave analysis on the available \pap and \NNbar cross sections~\cite{Zhou:2012ui}.\\
The Migdal-Watson approximation~\cite{MigdalWatson} is used as an approximate way to explicitly include the additional $j=X$ annihilation channels. This approximation relies on the fact that these $X$ multi-meson channels open below the \pap threshold and hence
the momentum dependence of the annihilation potential $V_{X\rightarrow p\overline p}$ around the \pap
threshold can be neglected.
The wave functions $\psi^{\rm PW}_{X\rightarrow p\overline p}$ for each PW can be rewritten in terms of the elastic component as $\omega_{\rm PW} \psi^{\rm PW}_{p\overline p\rightarrow p\overline p }$, with the weights $\omega_{\rm PW}$ to be determined from data. These latter weights are directly connected to the conversion weights $\omega_j$ in Eq.~(\ref{eq:corrfun_CC}) with the strong coupling term $\omega^s _j$ extended to the different PW states. A detailed estimate of the yields and kinematics of the annihilation channels ($\omega ^{\rm prod}_j$), necessary to isolate the strong coupling term in each PW, is not trivial since it involves contributions stemming from multi-pions channels and it should include also intermediate states of resonances strongly decaying into pions. For this reason, the extracted weights $\omega_{\rm PW}$ in this work contain informations not only on the coupling strength of the mesonic channels to \pap, but also on the abundances of the contributing multi-meson channels produced in the initial state.\\
The modeled correlation function reads~\cite{Haidenbauer:2018jvl}:

\begin{align}\label{eq:CFtotpApCC}
    C_{\pap}(\ks) & = \int S(r^*) |\psi_{p\overline p \rightarrow p\overline p}|^2 d^3r^* + \int S(r^*) |\psi_{n\overline n \rightarrow p\overline p}|^2 d^3r^*  +\sum_{\rm PW} \rho_{\rm PW} \omega_{\rm PW} \int S(r^*) |\psi^{\rm PW}_{ p\overline p \rightarrow p\overline p}|^2 d^3r^*\nonumber\\
    &= C_{p\overline p \rightarrow p\overline p}(\ks) + C_{n\overline n \rightarrow p\overline p}(\ks) + \sum_{\rm PW} C_{X \rightarrow p\overline p} ^{\rm PW}(\ks).
\end{align}

The first and second terms describe the elastic and \nan contributions, while the last term accounts for the annihilation channels. The degeneracy in spin and angular momentum is embedded in the statistical factors $\rho_{\rm PW}$.
To reduce the number of $\omega_{\rm PW}$ weights to be fitted, a study on the shape of the single inelastic correlation terms $C_{X \rightarrow p\overline p} ^{\rm PW}(\ks)$ is performed in each partial wave.
The correlations with a different profile in \ks are selected, allowing to determine three representative contributions: the \Swzero for S states, the \Pwunouno and \Pwtrezero for P states.\\
For the two systems containing strangeness, \paL and \LaL, no theoretical wave functions are currently available, hence the single-channel \Ledn analytical formula with a complex scattering length $f_0$ is used to evaluate the theoretical correlations~\cite{Lednicky:1981su,WUTPaper}. As mentioned above, in this single-channel approach, only the elastic contributions are explicitly accounted for in the \Rescatt, corresponding to the first term in Eq.~\ref{eq:corrfun_CC}. The imaginary part \Imscatt accounts for an average over all the inelastic contributions  of the \BBar interaction, mainly dominated by annihilation.
The same approach has been used in the ALICE femtoscopic measurements in \PbPb collisions~\cite{WUTPaper}, which delivered the only available scattering parameters on both the \paL and \LaL interaction. For the latter, theoretical predictions are available~\cite{BBar3}, providing values for the scattering parameters compatible with the ALICE \PbPb results~\cite{WUTPaper}.\\
The emitting source in Eq.~\ref{eq:corrfun_CC} can be determined as a function of the pair-transverse-mass \mt with a data-driven model based on proton-proton correlations~\cite{ALICE:Source}. This allows us to investigate the  interaction for different particle pairs.
The properties of the underlying interaction in \paL and \LaL systems do not depend on \mt and can hence be  better constrained using a \mt differential analysis. Considering the available sample, \num{6} and \num{3} \mt intervals are used for the \paL and \LaL measured correlations, respectively.
These experimental correlations are compared, in each \mt interval, to the \Ledn model by assuming at first the scattering parameters obtained in the \PbPb analysis~\cite{WUTPaper}. Secondly, a simultaneous fit for each pair in all the available \mt bins is performed leaving the \Imscatt to vary in order to test if a better agreement with the data is achieved. Further discussions on the two different fitting procedures can be found in the next section.\\

Experimentally, the correlation function is defined as

\begin{align}
  C(\ks)=\mathcal{N}\frac{N_\mathrm{SE}(\ks)}{N_\mathrm{ME}(\ks)}\xrightarrow{k^*\rightarrow\infty}1.  
\end{align}

Here $N_\mathrm{SE}(\ks)$ is the distribution of pairs measured in the same event, $N_\mathrm{ME}(\ks)$ is the reference distribution of uncorrelated pairs sampled from different (mixed) events and $\mathcal{N}$ is a normalization parameter determined by requiring that particle pairs with large \ks are not correlated. The mixed-event sample is obtained by pairing particles stemming from events with a similar number of charged particles at midrapidity and a close-by primary vertex position along the beam direction as done in ~\cite{ALICE:pSig0,ALICE:pK,ALICE:pOmega}.
The correlation functions of baryon--antibaryon and antibaryon--baryon pairs are combined to enhance the statistical significance for the \paL pairs, hence in the following \paL denotes the sum $\paL \oplus \apL$.
The \pap, \paL and \LaL data are fitted with a total correlation function

\begin{align}\label{eq:CFtot}
    C_{\rm{tot}}(\ks) = N_{D}\times C_{\mathrm{\rm{model}}}(\ks) \times C_{\rm{background}}(\ks),
\end{align}

where $N_D$ is a normalization constant fitted to data. The default fit range is $0<\ks<500$ \MeVc.
The modeled $C_{\rm{model}}(\ks) = 1 + \sum_{i} \lambda_{i} \times (C_{i}(\ks) - 1)$ includes the genuine ($i=\pap,\paL,\LaL$) correlation, estimated from Eq.~\ref{eq:CFtotpApCC} and using the \Ledn model, and the residual secondary contributions weighted by the $\lambda_i$ parameters~\cite{ALICE:Run1}.
The genuine contributions for \pap, \paL and \LaL amount to $\lambda_{\pap} = 66.5\%$, $\lambda_{\paL}=45.8\%$ and $\lambda_{\LaL}=30.9\%$, respectively. Residual contributions involving pairs measured in this work are modeled assuming the corresponding theoretical predictions mentioned above. Contributions involving $\Sigma^{\pm,0}$ ($\overline{\Sigma}^{\pm,0}$) and $\Xi^{-,0}$ ($\overline{\Xi}^{+,0}$) are considered to be constant in $k^*$ due to the limited theoretical knowledge, and amount to $10.1\%$, $44.6\%$ and $65.7\%$ for \pap, \paL and \LaL, respectively.
A crosscheck on these residuals by assuming a strong interaction based on the scattering parameters extracted in \PbPb measurements~\cite{WUTPaper} was performed and differences in the extracted results with respect to the constant assumption are found to be negligible.\\
A variation of $\pm10\%$ to the upper limit of the default fit range is applied for evaluating the systematic uncertainties. 
Additionally, the systematic uncertainties related to the $\lambda_i$ parameters are evaluated based on variations of the amount of secondary contributions to each measured particle species, where the largest source of uncertainty stems from the ratio $\Sigma^0$:$\,\Lambda = 0.33\pm0.07$~\cite{ALICE:Source,TFist,Albrecht:1986me,Sullivan:1987us,Yuldashev:1990az}. In addition to the feed-down contributions, a correction for finite experimental momentum resolution has to be taken into account for a direct comparison with data~\cite{ALICE:Run1}.\\
The size of the emitting source employed in the calculation of $C_{\mathrm{\rm{model}}}(\ks)$ for the three \BBar pairs is fixed from the data-driven analysis of \pprot pairs, which demonstrates the existence of a common Gaussian core as a function of \mt
for all baryon--baryon pairs when contributions from short-lived strongly decaying resonances are properly included~\cite{ALICE:Source}.
For the \pap pairs, the core source size at the corresponding $\left<m_\mathrm{T}\right>=1.45\,\,\mathrm{Ge\kern-.2emV/}c^2$ is $r_\mathrm{core}=1.06\pm0.04~$ fm and the associated effective Gaussian source size is $r_0=1.22$ fm. The core radii for the \paL and \LaL \mt bins presented in this letter are $r_\mathrm{core}(\left<m_\mathrm{T}\right>=1.75 \,\,\mathrm{Ge\kern-.2emV/}c^2)=0.95\pm0.04~$ fm ($r_0 =1.15$ fm) and $r_\mathrm{core}(\left<m_\mathrm{T}\right>=2.12\,\, \mathrm{Ge\kern-.2emV/}c^2)=0.87\pm0.04~$ fm ($r_0 =1.11$ fm), respectively.\\
The second term in Eq.~\ref{eq:CFtot}, $C_{\rm{background}}(\ks)$,  accounts for non-femtoscopic effects due to energy-momentum conservation at large \ks~\cite{ALICE:Run1} and to minijet phenomena arising from hard processes at the parton level, largely present in the measurement of \BBar correlations:

\begin{align}\label{eq:totbackground}
C_{\rm{background}} (\ks) & = C_{\rm{minijet}}(\ks) + C_{\rm{baseline}}(\ks) \nonumber \\
& = [w_C C_C (\ks)+(1-w_C)C_{NC} (\ks)]+(a+b\ks).
\end{align}

\begin{figure}[h!]
    \centering
    \includegraphics[width=0.495 \textwidth]{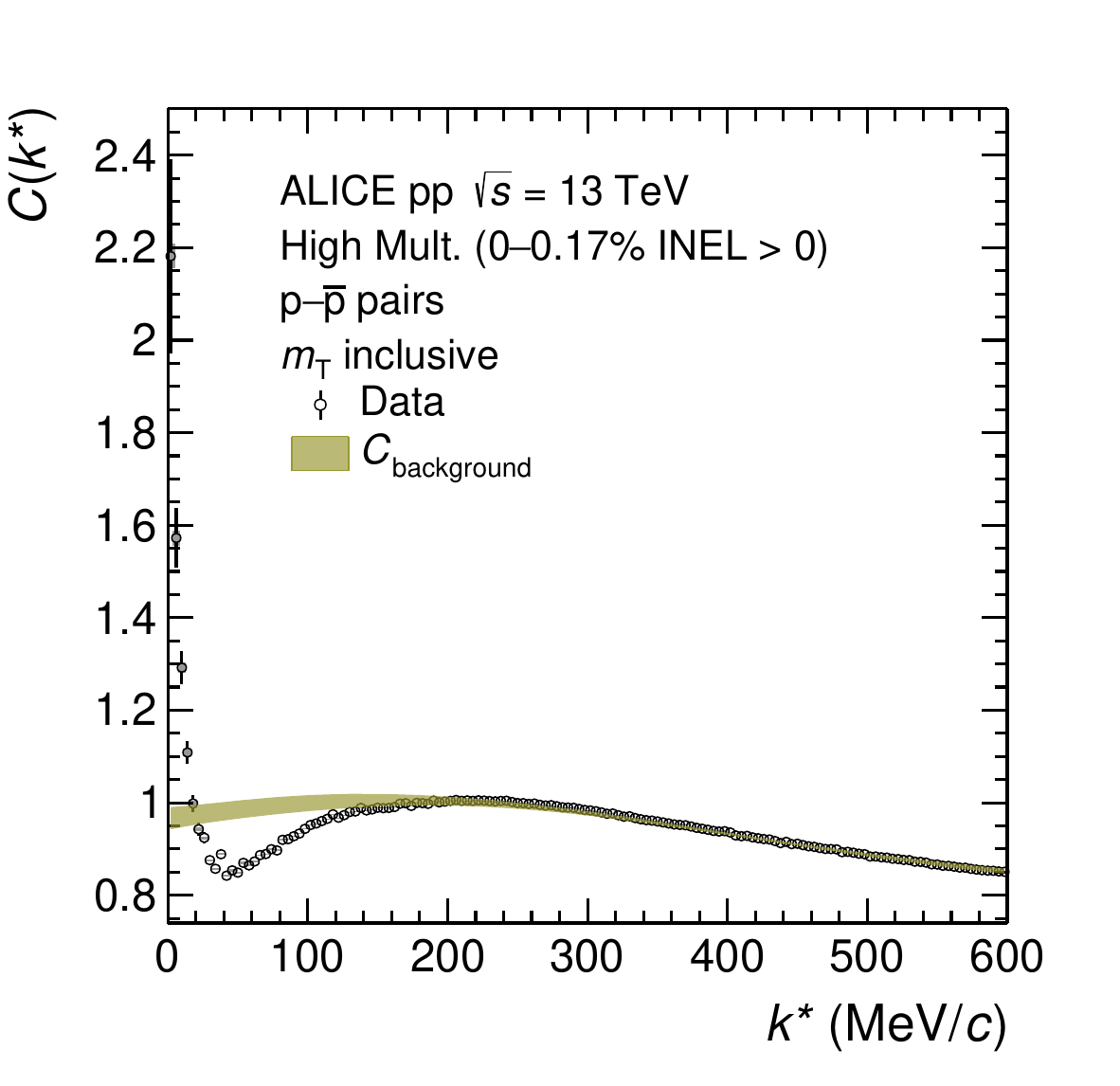}
    \caption{(Color online) 
     Measured \pap correlation function (empty points) with statistical (line) and systematic (grey boxes) uncertainties. The band represents the $C_{\rm{background}}(\ks)$ fit as described in the text.
}
    \label{fig:Minijet}
\end{figure}

A data-driven approach is employed using PYTHIA 8.2~\cite{PYTHIA} to model the mini-jet part contained in $C_{\rm{background}}(\ks)$.
The particle production in such simulations is associated to two processes: particles stemming from a common parton (common ancestors), leading to the minijet component, and particles coming from different partons (non-common ancestors), responsible for the non-jet part. 
The $C_{\rm{minijet}}(\ks)$ in Eq.~(\ref{eq:totbackground}) is given by a linear combination of the common ($C_C (\ks)$) and non-common ($C_{NC}(\ks)$) contributions weighted by a factor $w_C$ and $(1-w_{C})$, respectively. The ancestor weight $w_C$ is a free parameter in the fit of $C_{\rm{tot}}(\ks)$ to the data.
The common and non-common correlations obtained from PYTHIA 8.2 are fitted with a product of three Gaussian functions up to $\ks=2500$ \MeVc, providing good agreement with the simulated data.
To account for remaining non-femtoscopic effects at large \ks~\cite{ALICE:Run1}, a linear baseline $C_{\rm{baseline}}(\ks)=a+b\ks$ is added to the ancestors term in Eq.~(\ref{eq:totbackground}).
The coefficients $a$ and $b$ are fixed by fitting $C_{\rm{background}}(\ks)$ to the data in the region of $400 < \ks<2500$ \MeVc. The results for \pap pairs are shown in Fig.~\ref{fig:Minijet}. The band represents the $1\sigma$ uncertainty associated to the template fitting.
The shape of $C_{\rm{background}}(\ks)$ agrees within uncertainties with the data in the region above $\ks \approx 200$ \MeVc, where the non-flat behavior of minijet contributions is visible. A change of $\pm10\%$ in this range and a quadratic polynomial are included to estimate the systematic uncertainty related to the total background. Similar results and conclusions are obtained for the \paL and \LaL systems. 

\section{Results}\label{sec:Results}
\begin{figure}[h!]
    \centering
    \includegraphics[width=0.495 \textwidth]{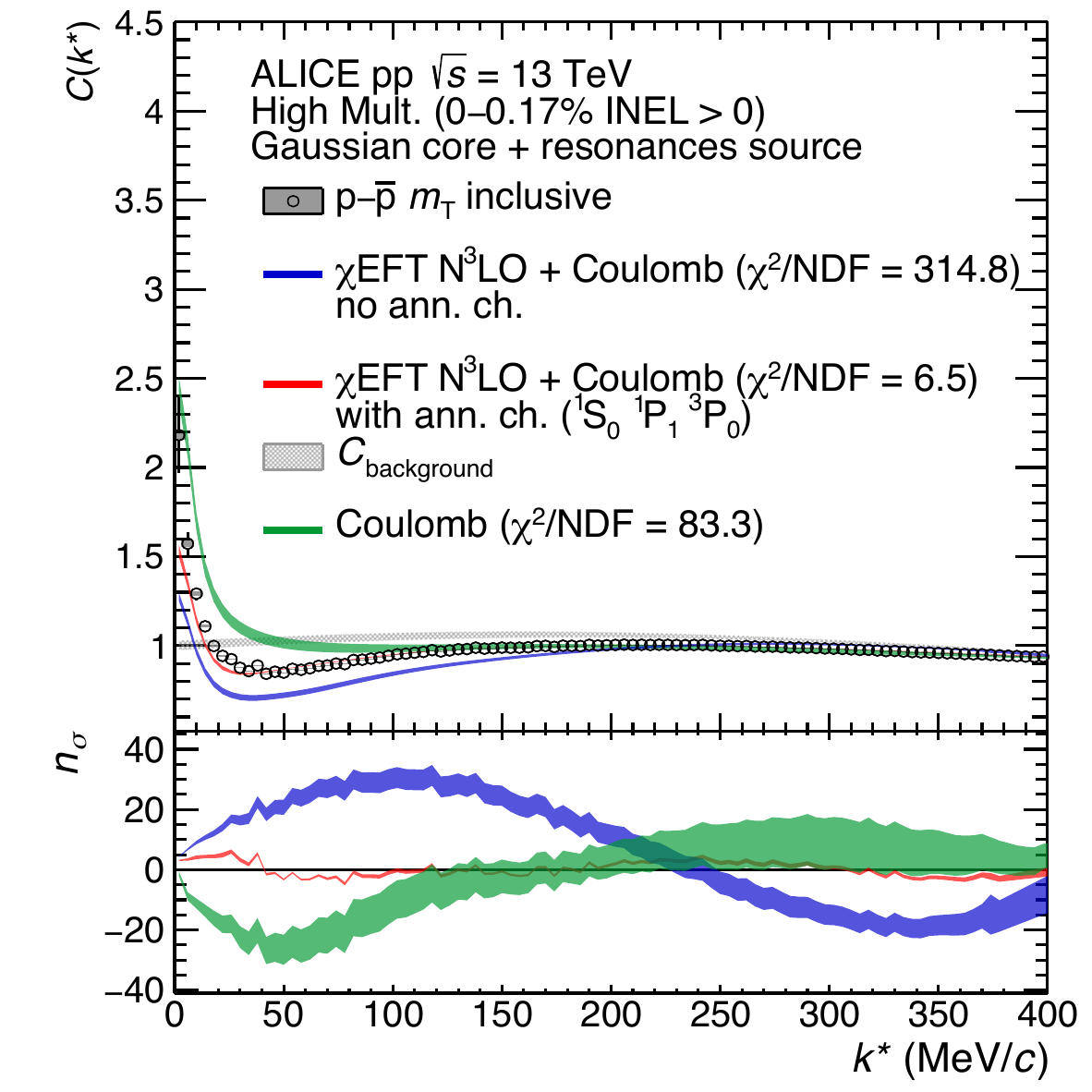}
    \caption{(Color online) Measured correlation function of \pap pairs. Statistical (bars) and systematic (boxes) uncertainties are shown separately. The Coulomb only interaction is shown by the green band. The blue band represents the fit performed using $\rm N^3$LO \chiEFT potentials~\cite{Dai:2017ont} with elastic and \nan coupled-channel. The inclusion of annihilation channels is shown by the red band, along with the $C_{\rm{background}}(\ks)$, multiplied by the normalization constant $N_D$ obtained in the fit. The reported average $\chi^2/\rm NDF$ is evaluated in the \ks interval $[0,400]$ \MeVc and it includes correlations between the data points.
    Lower panel: $n_\sigma$ deviation between data and model in terms of numbers of statistical standard deviations.}
    \label{fig:pApCFs}
\end{figure}

\begin{figure}[h!]
    \centering
    \includegraphics[width=0.495 \textwidth]{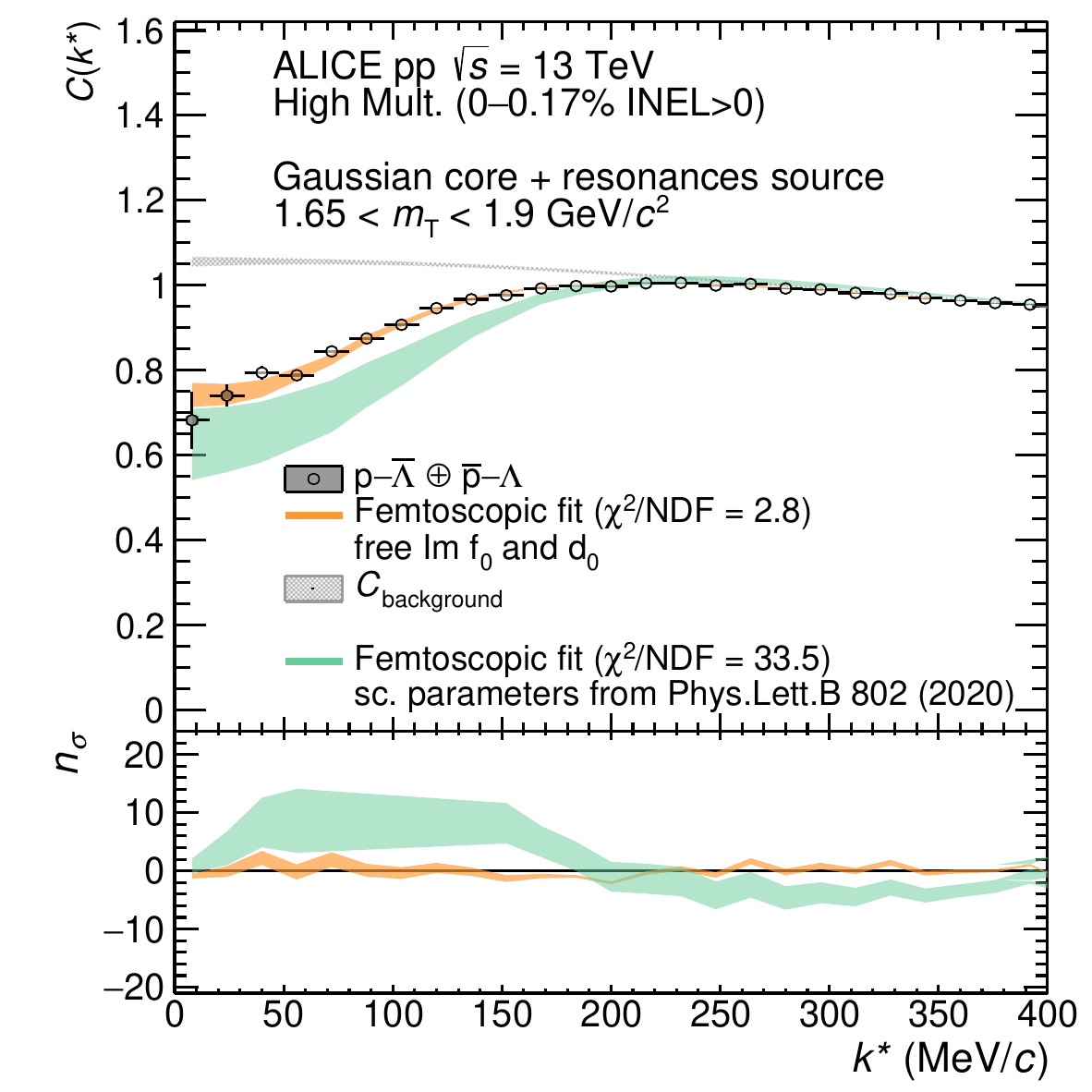}
    \includegraphics[width=0.495 \textwidth]{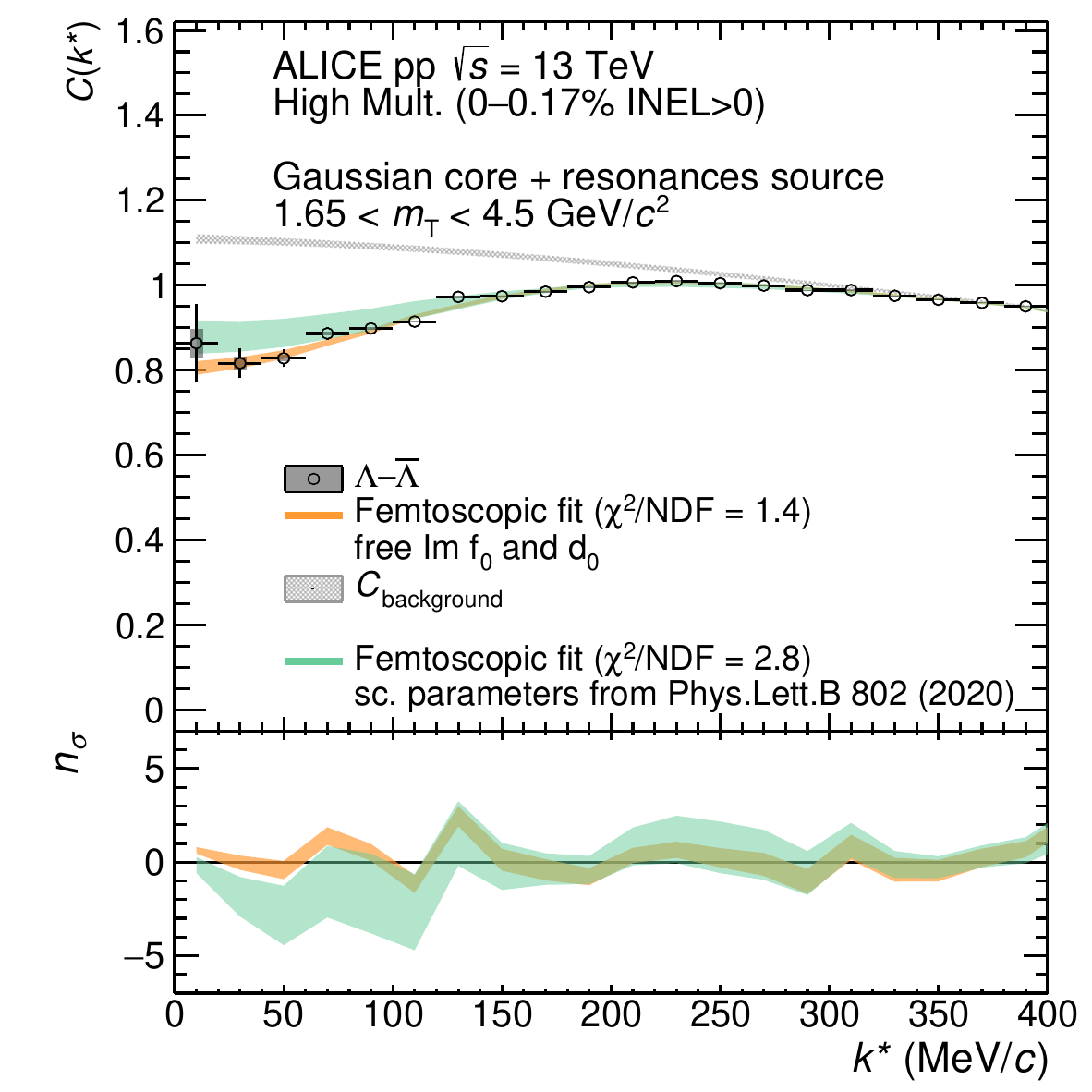}
    \caption{(Color online) Measured correlation function of \paL (left) and \LaL (right) pairs for two representative \mt bins. Statistical (bars) and systematic (boxes) uncertainties are shown separately. Results using the \Ledn formula with \PbPb scattering parameters~\cite{WUTPaper} are shown in light green. Orange bands are the results with \effran and \Imscatt as free parameters. In grey the corresponding $C_{\rm{background}}(\ks)$, multiplied by the normalization constant $N_D$, is shown. The reported average $\chi^2/\rm NDF$ is evaluated in the \ks interval $[0,400]$ \MeVc and it includes correlations between the data points.
    Lower panel: same as in Fig.~\ref{fig:pApCFs}.}
    \label{fig:otherCFs}
\end{figure}

\noindent
The correlation functions for \pap and for two representative \mt bins of \paL and \LaL are shown in Fig.~\ref{fig:pApCFs} and in Fig.~\ref{fig:otherCFs}, respectively. The results for the remaining \mt bins are presented in Figs.~\ref{fig:otherCFs_pALmT},~\ref{fig:otherCFs_LALmT} of Appendix~\ref{App:pALLAL}.
The lower panels show the statistical deviation between data and model expressed in terms of numbers of standard deviation $n_\sigma$. The width of the band represents the total uncertainty of the fit. The grey boxes correspond to the systematic uncertainties of the data. They are maximal at the lowest \ks bin and amount to $1\%$, $4\%$ and $10\%$ for \pap, \paL and \LaL pairs, respectively. 
The $C_{\pap}(\ks)$ correlation is compared first to a Coulomb-only interaction and secondly to a Coulomb + strong interaction from \NNbar \chiEFT potentials with wave functions for the $\nan \rightarrow \pap$ process explicitly included~\cite{Dai:2017ont}.
Results for this latter scenario are obtained by evaluating the genuine \pap correlation in Eq.~(\ref{eq:CFtotpApCC}) with only the first two terms and shown in blue in Fig.~\ref{fig:pApCFs}. 
The opening of the \nan channel above threshold, expected as a cusp structure in the \CF at $\ks\approx 50$ \MeVc, is not visible in agreement with the weak coupling already measured in scattering experiments~\cite{Zhou:2012ui}.
The chiral model underestimates the data in the region below 200 \MeVc and it cannot reproduce the enhancement above unity of the \CF as \ks approaches zero. 
This increase is not described either by assuming only the Coulomb attraction (green band), showing that annihilation is largely present close to threshold as $\ks \rightarrow 0$ \MeVc.
The contributions to the \pap correlation from the multi-meson annihilation channels, produced as initial states which feed into the measured \pap system, are not explicitly accounted for in the chiral potential and hence the last term $\sum_{PW} C_{X \rightarrow p\overline p} ^{PW}(\ks)$ in Eq.(\ref{eq:CFtotpApCC}) is currently missing in the fit shown by the blue band.\\
The red band in Fig.~\ref{fig:pApCFs} represents the results obtained from the explicit inclusion of the annihilation channels in the third term of Eq.~(\ref{eq:CFtotpApCC}) via the Migdal-Watson approximation.
The corresponding fit provides a better description of the data in the low \ks region where annihilation is dominant.
The extracted coupling weights $\omega_{\rm PW}$ from this femtoscopic fit are \SwzeroResult and \PwtrezeroResult, while $\omega_{\Pwunouno}$ is compatible with zero. 
The hierarchy of the coupling weights in the different PW agrees with the inelasticity parameters $\eta$ obtained in the recent partial-wave analysis~\cite{Zhou:2012ui}.

For the systems containing strangeness, the Migdal-Watson approach cannot be employed since only scattering parameters for the \paL and \LaL interaction are available~\cite{WUTPaper}. 
The values of \Rescatt and \Imscatt obtained in \PbPb measurements are employed in the \Ledn analytical formula~\cite{Lednicky:1981su,WUTPaper} to model the \paL and \LaL genuine correlation functions.
In Fig.~\ref{fig:otherCFs}, the results obtained modeling the \paL and \LaL theoretical correlations with the \Ledn model described in Sec.~\ref{sec:AnalysisCF} are shown.
The first tested scenario assumes the scattering parameters extracted from \PbPb results~\cite{WUTPaper} and the results are denoted by light green bands. 
It can be expected that if the direct contributions of the annihilation channels are negligible, the values extracted in \PbPb will reproduce well also the \pp data in this analysis.
As can be seen from the right panel in Fig.~\ref{fig:otherCFs}, this first approach reproduces the measured \LaL correlation function, with an average $\chi^2/\rm NDF =2.8$ evaluated in the \ks interval $[0,400]$ \MeVc but it clearly underestimates the \paL correlation data in the \ks region below $200$ \MeVc. 
A similar trend is observed when performing the fit to the \pap measured correlation with the \Ledn approach used in the \PbPb results~\cite{WUTPaper} as shown in Fig.~\ref{fig:pAp_Lednicky} in the Appendix~\ref{App:pAp}. The discrepancy hence, as in the \pap case, has to be attributed to a larger amount of annihilation channels feeding into the \paL system with respect to the \LaL pairs.
To validate this interpretation, a simultaneous fit in all the \mt bins is performed leaving free to vary the imaginary part of the scattering length \Imscatt, accounting for inelastic channels, and the effective range \effran. The negative real part of the scattering length \Rescatt, indicating either a repulsive elastic interaction or a possible bound state, is kept fixed to the \PbPb results~\cite{WUTPaper}. 
To reach a reasonable agreement of the model with \paL data, \Imscatt has to be increased by approximately a factor 5.3, while the change in the extracted \effran is negligible. Such a discrepancy can be attributed to the failure of the single-channel \Ledn model to properly accommodate the direct contribution of inelastic channels (last term in Eq.~(\ref{eq:corrfun_CC})). A similar fit is applied to the \LaL system and values of \Imscatt and \effran compatible with the \PbPb measurements are found, implying a negligible effect of the direct contribution of annihilation channels. The corresponding results are shown in Fig~\ref{fig:otherCFs} (orange band), for \paL (left panel) and \LaL (right panel). A similar trend is obtained in the remaining \mt intervals and shown in Appendix~\ref{App:pALLAL}.\\
The different results for the \paL and \LaL systems may also be related to a different amount of initially produced multi-meson states feeding into the two \BBar pairs. 
To substantiate this scenario, a study of the two-meson channel contributions ($\uppi\overline{\uppi}$, $\uppi\overline{\rm K}$) is performed using the EPOS transport model~\cite{Pierog:2013ria}.
The fraction of two--mesons pairs $f_{2M\rightarrow B \overline B}$ produced in the initial collision and kinematically available to produce \BBar pairs with low \ks is estimated.
The latter is obtained by dividing the amount of meson--meson pairs initially produced, having a center-of-mass energy above the \BBar threshold and leading to \BBar pairs at low \ks, by the total number of produced two--meson pairs kinematically allowed to create the \BBar pairs.
Based on this study and considering these kinematics considerations, 
a similar amount ($\approx 6.4\%$) is found for \paL and \LaL pairs,  indicating that the above effect is related to the properties of the \paL and \LaL interaction. To quantify the final relative amount of annihilation channels feeding to the \paL and \LaL systems, the fractions have to be multiplied by the corresponding coupling constant $g$, obtained within an $\rm SU(3)$ Lagrangian by evaluating the trace of the meson--baryon interaction term~\cite{Giacosa}. Within this simplified calculations, the coupling strength for the \paL system is found to be approximately 3.3 times larger than for the \LaL pairs.\\
The estimated contribution $g\times f_{2M\rightarrow B \overline B}$, although limited to only two-meson channels, for \paL pairs is found to be about 6 times larger than for \LaL pairs, indicating a different annihilation contributions occurring in \paL and \LaL interaction which is confirmed by the measured correlation functions in Fig.~\ref{fig:otherCFs}. 
These estimations, even though based on a qualitative approach, clearly indicates that the annihilation for the \LaL interaction is present but it should not be largely dominant over the elastic part. More input from theory is needed in order to claim if such a condition is ideal for the formation of bound-states in the \LaL system.
The results for the \paL system, however, clearly point to a much larger presence of the annihilation channels, which might reduce the possibility to create baryonia.
The data presented in Fig.~\ref{fig:otherCFs} represent the most precise data currently available on \paL and \LaL pairs and can provide constraints for theoretical models on these interactions.

In conclusion, femtoscopic techniques have been adopted to study the annihilation dynamics in \pap, \paL and \LaL systems.
A quantitative determination of the effective coupling weights, connected to the annihilation channels present in \pap, has been obtained adopting a coupled-channel approach with $\rm N^3$LO \chiEFT potentials~\cite{Dai:2017ont}. The largest couplings have been obtained in the spin triplet P (\Pwtrezero) and singlet S (\Swzero) state. The inclusion of these inelastic channels leads to a better agreement between data and model in the region of \ks below 50 \MeVc, indicating a wide presence of annihilation channels close to threshold.
The scattering parameters obtained in \PbPb collisions~\cite{WUTPaper} have been used to model the \paL and \LaL data using the \Ledn formula. 
A consistent description of the \LaL correlation is achieved while an increase of the \Imscatt in the \paL interaction is needed to improve the agreement with the \paL data. These results, confirmed by kinematics and $\rm SU(3)$ flavor symmetry considerations, indicate a larger contribution in \paL from annihilation channels in comparison to \LaL.
The ALICE data shown in this work delivered the most precise measurements on \pap, \paL and \LaL systems at low momenta and suggest that baryonia are unlikely to occur in \pap and \paL systems due to the large annihilation contributions present for these pairs. A modeling of the \BBar interaction for systems as \paL, based on optical potentials, and a quantitative estimate of production of the multi-meson annihilation channels in the collisions, can provide a better understanding of the elastic and the annihilation term which can help to strengthen final conclusions on possible bound states.

\newenvironment{acknowledgement}{\relax}{\relax}
\begin{acknowledgement}
\section*{Acknowledgements}
The ALICE Collaboration is grateful to Prof.~Johann Haidenbauer and Prof.~Francesco Giacosa for the extremely valuable guidance on the theoretical aspects and fruitful discussions.

The ALICE Collaboration would like to thank all its engineers and technicians for their invaluable contributions to the construction of the experiment and the CERN accelerator teams for the outstanding performance of the LHC complex.
The ALICE Collaboration gratefully acknowledges the resources and support provided by all Grid centres and the Worldwide LHC Computing Grid (WLCG) collaboration.
The ALICE Collaboration acknowledges the following funding agencies for their support in building and running the ALICE detector:
A. I. Alikhanyan National Science Laboratory (Yerevan Physics Institute) Foundation (ANSL), State Committee of Science and World Federation of Scientists (WFS), Armenia;
Austrian Academy of Sciences, Austrian Science Fund (FWF): [M 2467-N36] and Nationalstiftung f\"{u}r Forschung, Technologie und Entwicklung, Austria;
Ministry of Communications and High Technologies, National Nuclear Research Center, Azerbaijan;
Conselho Nacional de Desenvolvimento Cient\'{\i}fico e Tecnol\'{o}gico (CNPq), Financiadora de Estudos e Projetos (Finep), Funda\c{c}\~{a}o de Amparo \`{a} Pesquisa do Estado de S\~{a}o Paulo (FAPESP) and Universidade Federal do Rio Grande do Sul (UFRGS), Brazil;
Ministry of Education of China (MOEC) , Ministry of Science \& Technology of China (MSTC) and National Natural Science Foundation of China (NSFC), China;
Ministry of Science and Education and Croatian Science Foundation, Croatia;
Centro de Aplicaciones Tecnol\'{o}gicas y Desarrollo Nuclear (CEADEN), Cubaenerg\'{\i}a, Cuba;
Ministry of Education, Youth and Sports of the Czech Republic, Czech Republic;
The Danish Council for Independent Research | Natural Sciences, the VILLUM FONDEN and Danish National Research Foundation (DNRF), Denmark;
Helsinki Institute of Physics (HIP), Finland;
Commissariat \`{a} l'Energie Atomique (CEA) and Institut National de Physique Nucl\'{e}aire et de Physique des Particules (IN2P3) and Centre National de la Recherche Scientifique (CNRS), France;
Bundesministerium f\"{u}r Bildung und Forschung (BMBF) and GSI Helmholtzzentrum f\"{u}r Schwerionenforschung GmbH, Germany;
General Secretariat for Research and Technology, Ministry of Education, Research and Religions, Greece;
National Research, Development and Innovation Office, Hungary;
Department of Atomic Energy Government of India (DAE), Department of Science and Technology, Government of India (DST), University Grants Commission, Government of India (UGC) and Council of Scientific and Industrial Research (CSIR), India;
Indonesian Institute of Science, Indonesia;
Istituto Nazionale di Fisica Nucleare (INFN), Italy;
Institute for Innovative Science and Technology , Nagasaki Institute of Applied Science (IIST), Japanese Ministry of Education, Culture, Sports, Science and Technology (MEXT) and Japan Society for the Promotion of Science (JSPS) KAKENHI, Japan;
Consejo Nacional de Ciencia (CONACYT) y Tecnolog\'{i}a, through Fondo de Cooperaci\'{o}n Internacional en Ciencia y Tecnolog\'{i}a (FONCICYT) and Direcci\'{o}n General de Asuntos del Personal Academico (DGAPA), Mexico;
Nederlandse Organisatie voor Wetenschappelijk Onderzoek (NWO), Netherlands;
The Research Council of Norway, Norway;
Commission on Science and Technology for Sustainable Development in the South (COMSATS), Pakistan;
Pontificia Universidad Cat\'{o}lica del Per\'{u}, Peru;
Ministry of Education and Science, National Science Centre and WUT ID-UB, Poland;
Korea Institute of Science and Technology Information and National Research Foundation of Korea (NRF), Republic of Korea;
Ministry of Education and Scientific Research, Institute of Atomic Physics and Ministry of Research and Innovation and Institute of Atomic Physics, Romania;
Joint Institute for Nuclear Research (JINR), Ministry of Education and Science of the Russian Federation, National Research Centre Kurchatov Institute, Russian Science Foundation and Russian Foundation for Basic Research, Russia;
Ministry of Education, Science, Research and Sport of the Slovak Republic, Slovakia;
National Research Foundation of South Africa, South Africa;
Swedish Research Council (VR) and Knut \& Alice Wallenberg Foundation (KAW), Sweden;
European Organization for Nuclear Research, Switzerland;
Suranaree University of Technology (SUT), National Science and Technology Development Agency (NSDTA) and Office of the Higher Education Commission under NRU project of Thailand, Thailand;
Turkish Energy, Nuclear and Mineral Research Agency (TENMAK), Turkey;
National Academy of  Sciences of Ukraine, Ukraine;
Science and Technology Facilities Council (STFC), United Kingdom;
National Science Foundation of the United States of America (NSF) and United States Department of Energy, Office of Nuclear Physics (DOE NP), United States of America.

In addition, individual groups and members have received support from Horizon 2020 and Marie Skłodowska Curie Actions, European Union.

\end{acknowledgement}

\bibliographystyle{utphys}   
\bibliography{bibliography}
\newpage
\appendix

\section{Additional material}
\subsection{\paL and \LaL results in \mt intervals}\label{App:pALLAL}
\begin{figure}[h!]
    \centering
    \includegraphics[width=0.43 \textwidth]{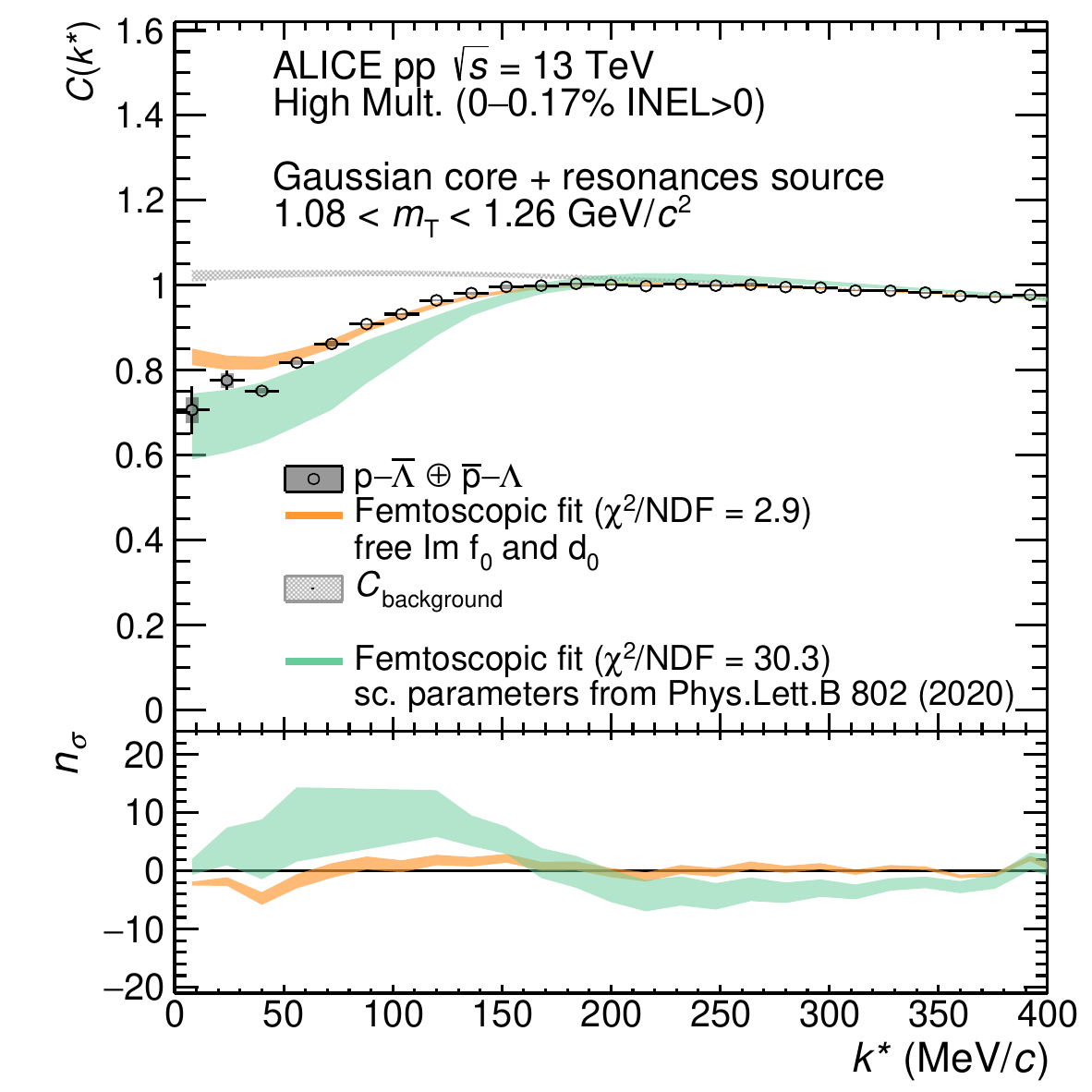}
    \includegraphics[width=0.43 \textwidth]{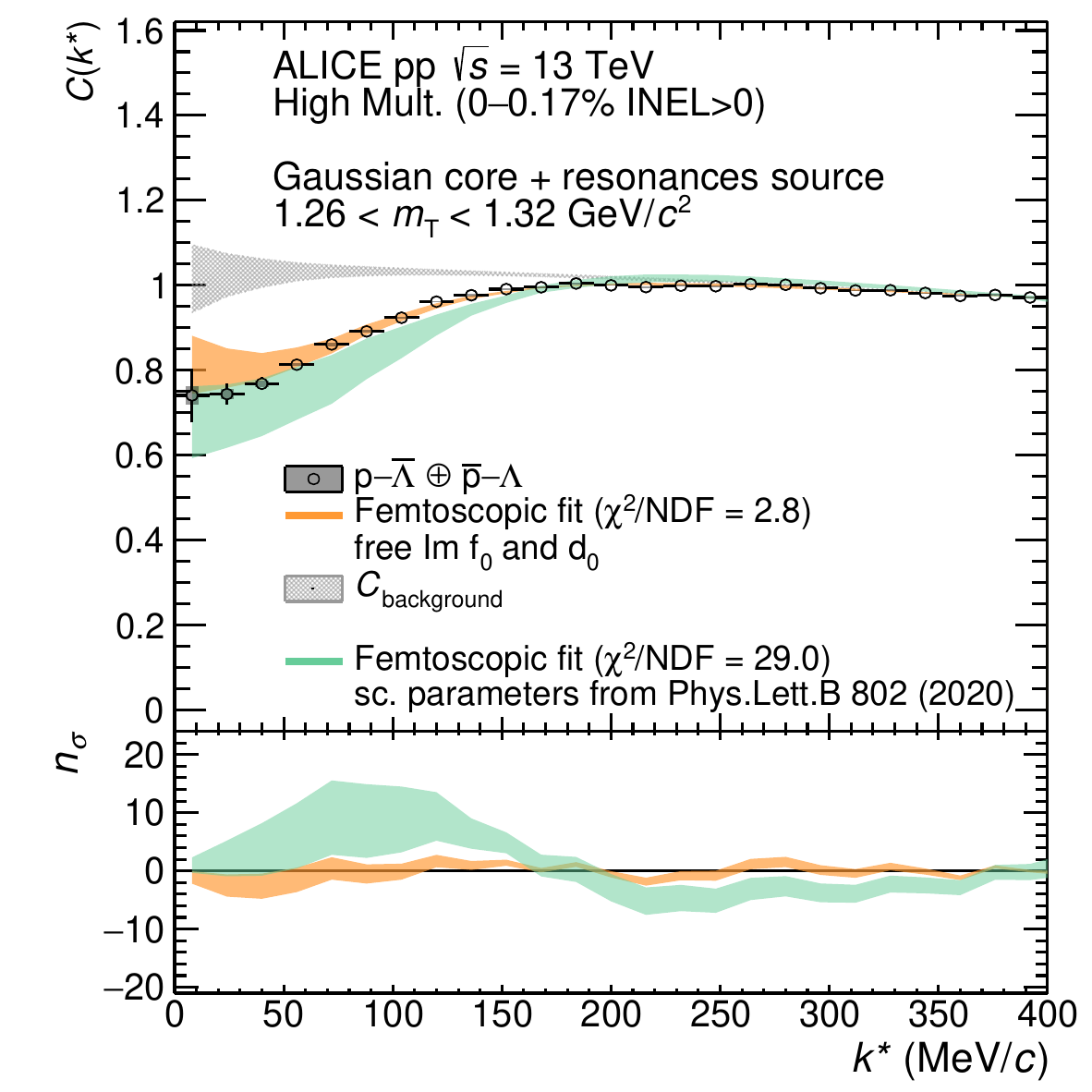}\\
    \includegraphics[width=0.43 \textwidth]{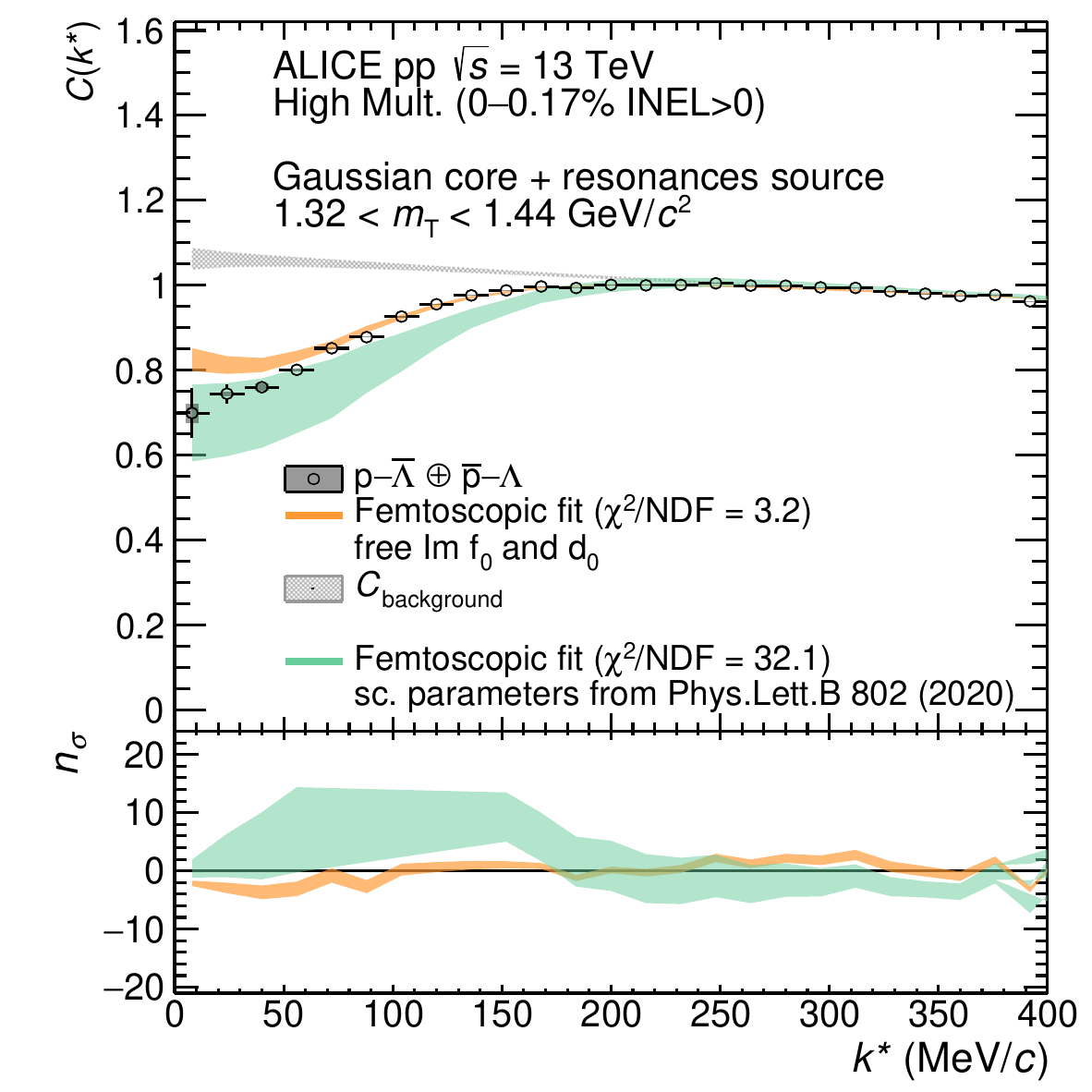}
    \includegraphics[width=0.43 \textwidth]{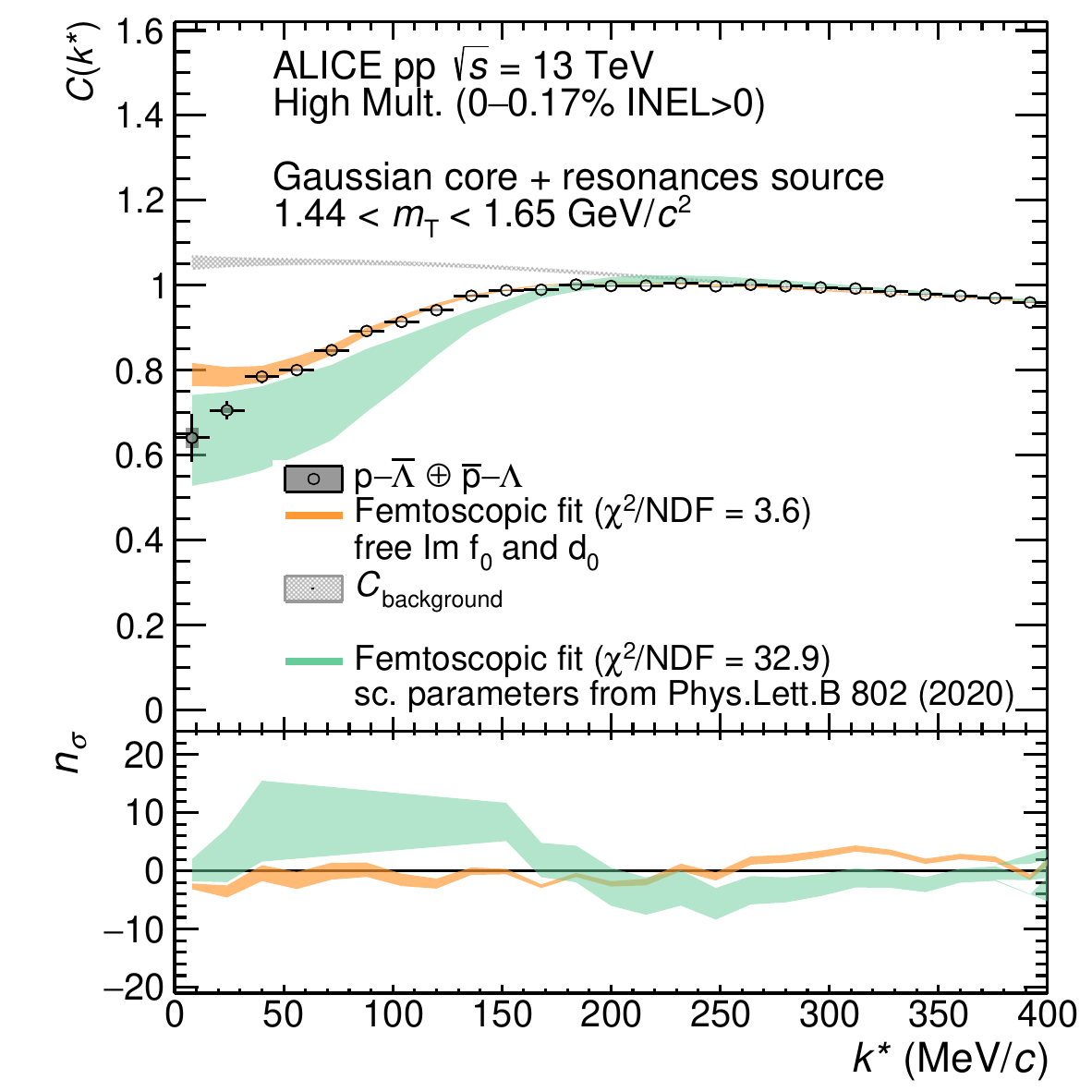}\\
    \includegraphics[width=0.43 \textwidth]{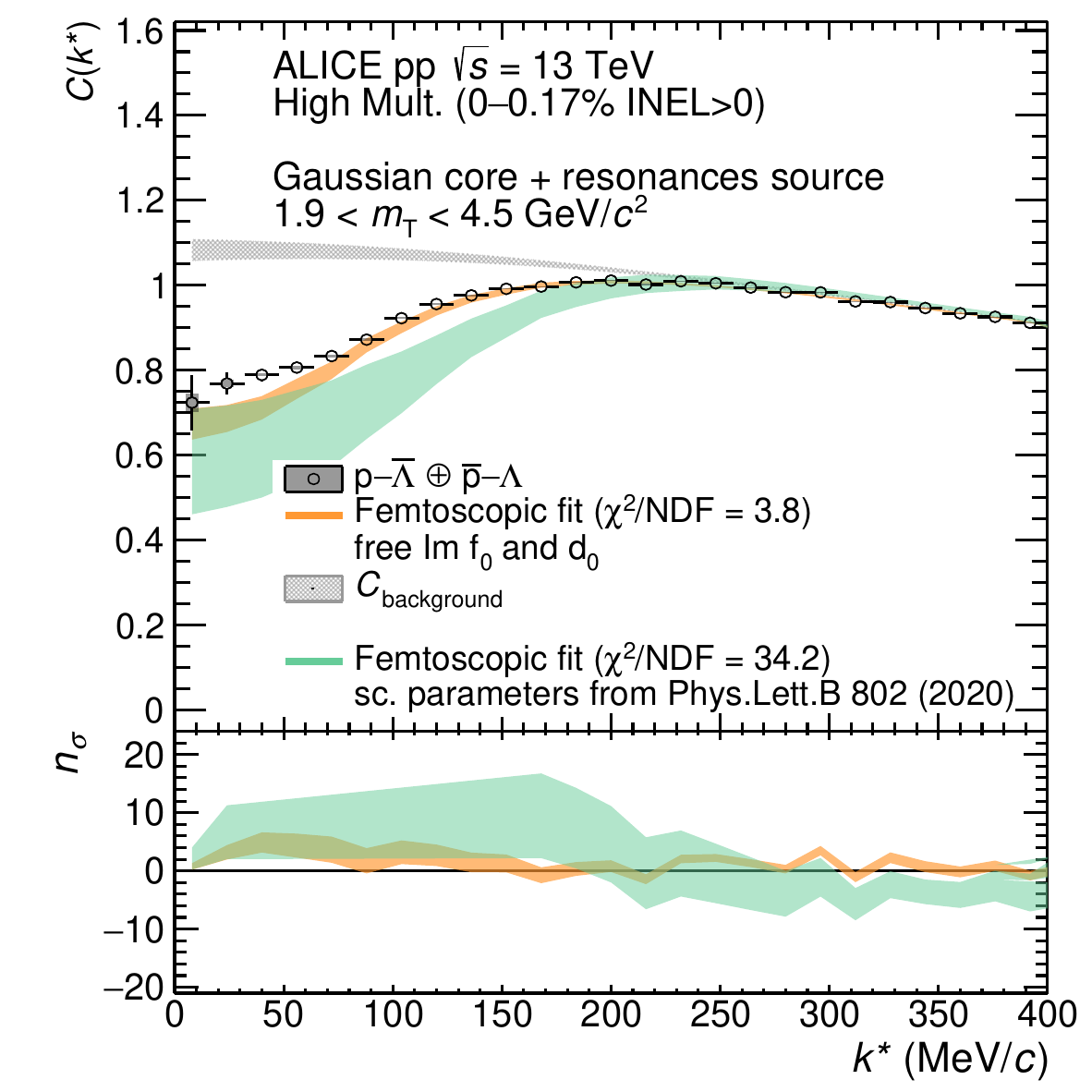}
    \caption{(Color online) Measured correlation function of \paL pairs for remaining \mt bins.
    Same description as in Fig.~\ref{fig:otherCFs}.}
    \label{fig:otherCFs_pALmT}
\end{figure}

\begin{figure}[h!]
    \centering
    \includegraphics[width=0.43 \textwidth]{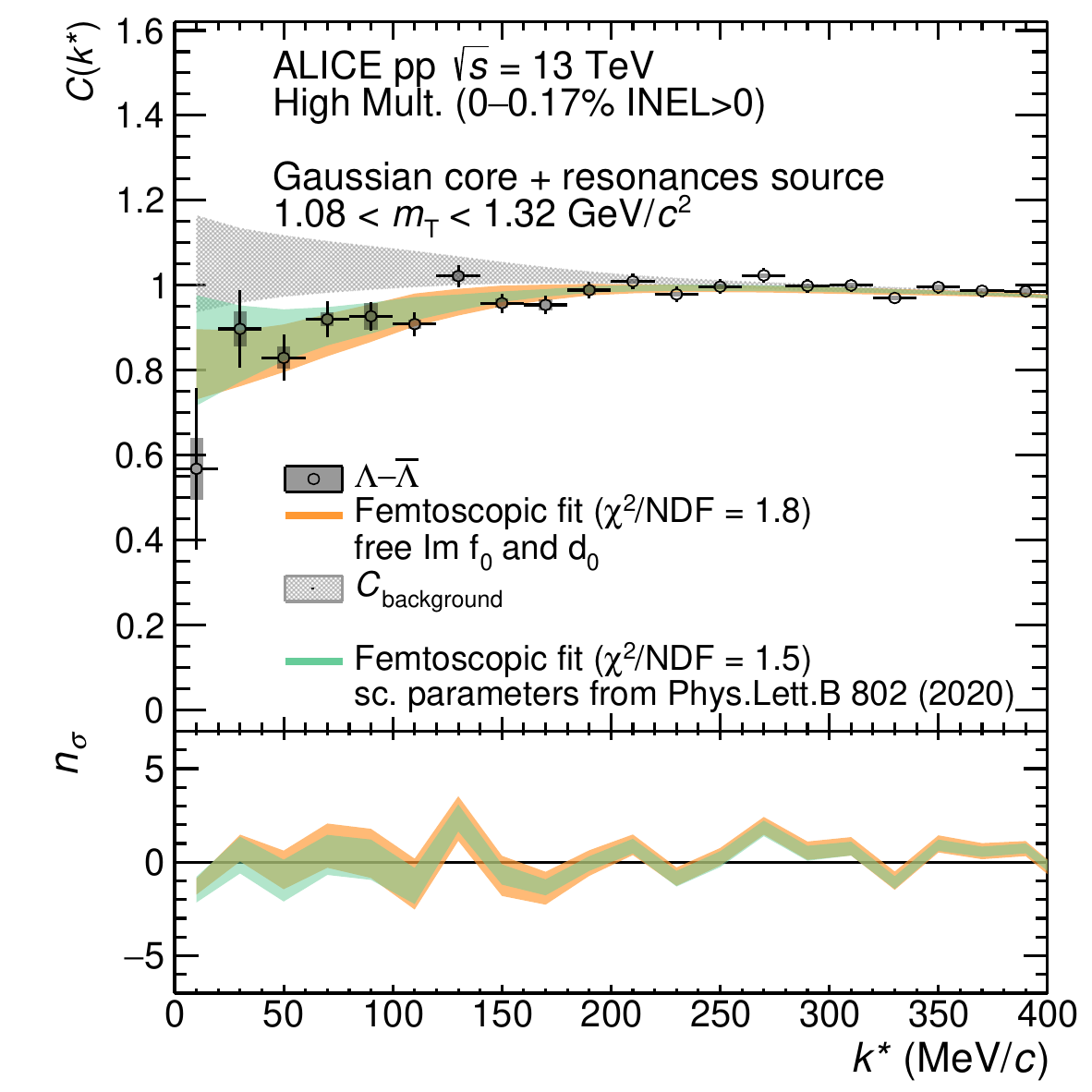}
    \includegraphics[width=0.43 \textwidth]{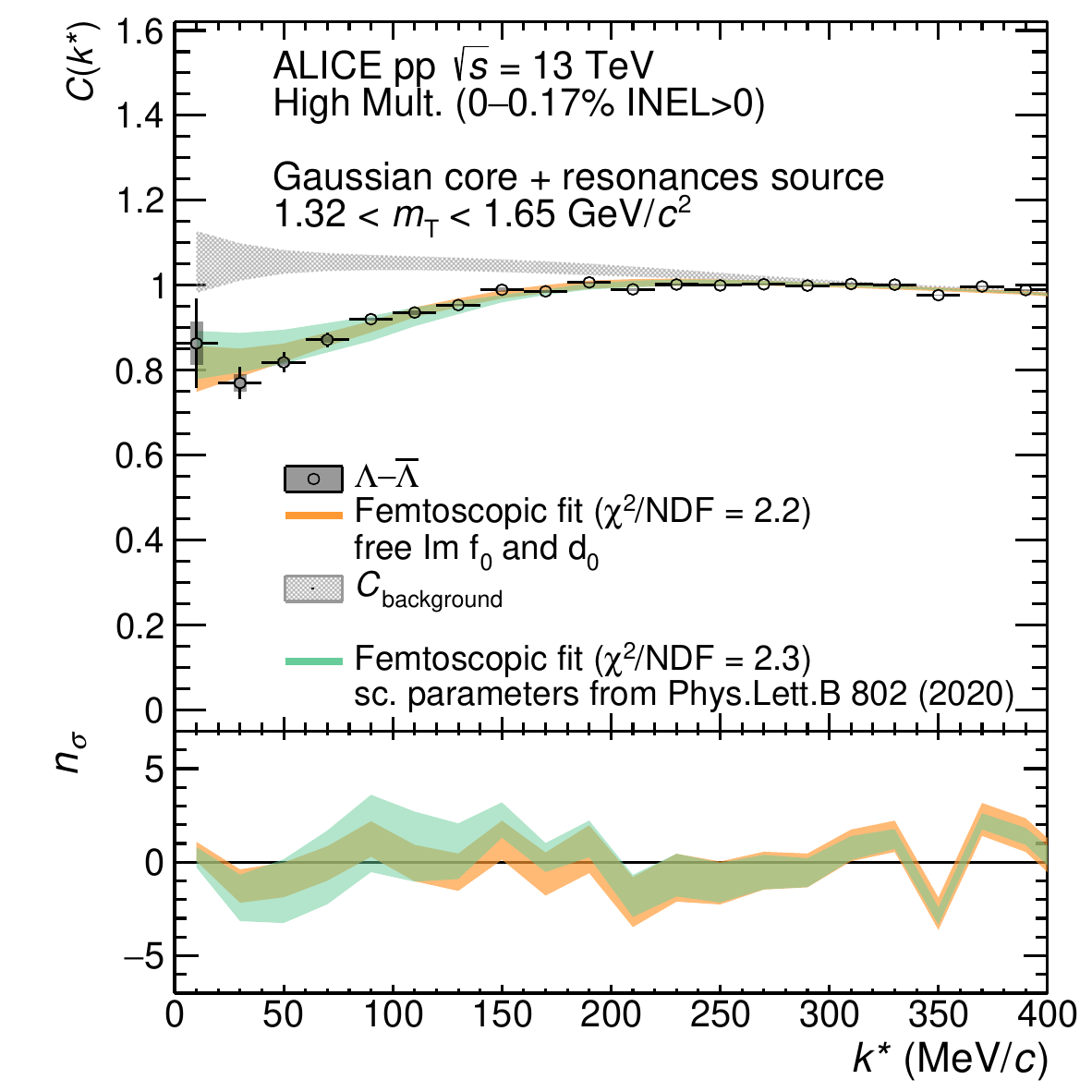}\\
    \caption{(Color online) Measured correlation function of \LaL pairs for remaining \mt bins.
    Same description as in Fig.~\ref{fig:otherCFs}.}
    \label{fig:otherCFs_LALmT}
\end{figure}
\newpage
\subsection{Results on \pap pairs with the \Ledn model}\label{App:pAp}
\begin{figure}[h!]
    \centering
    \includegraphics[width=0.45 \textwidth]{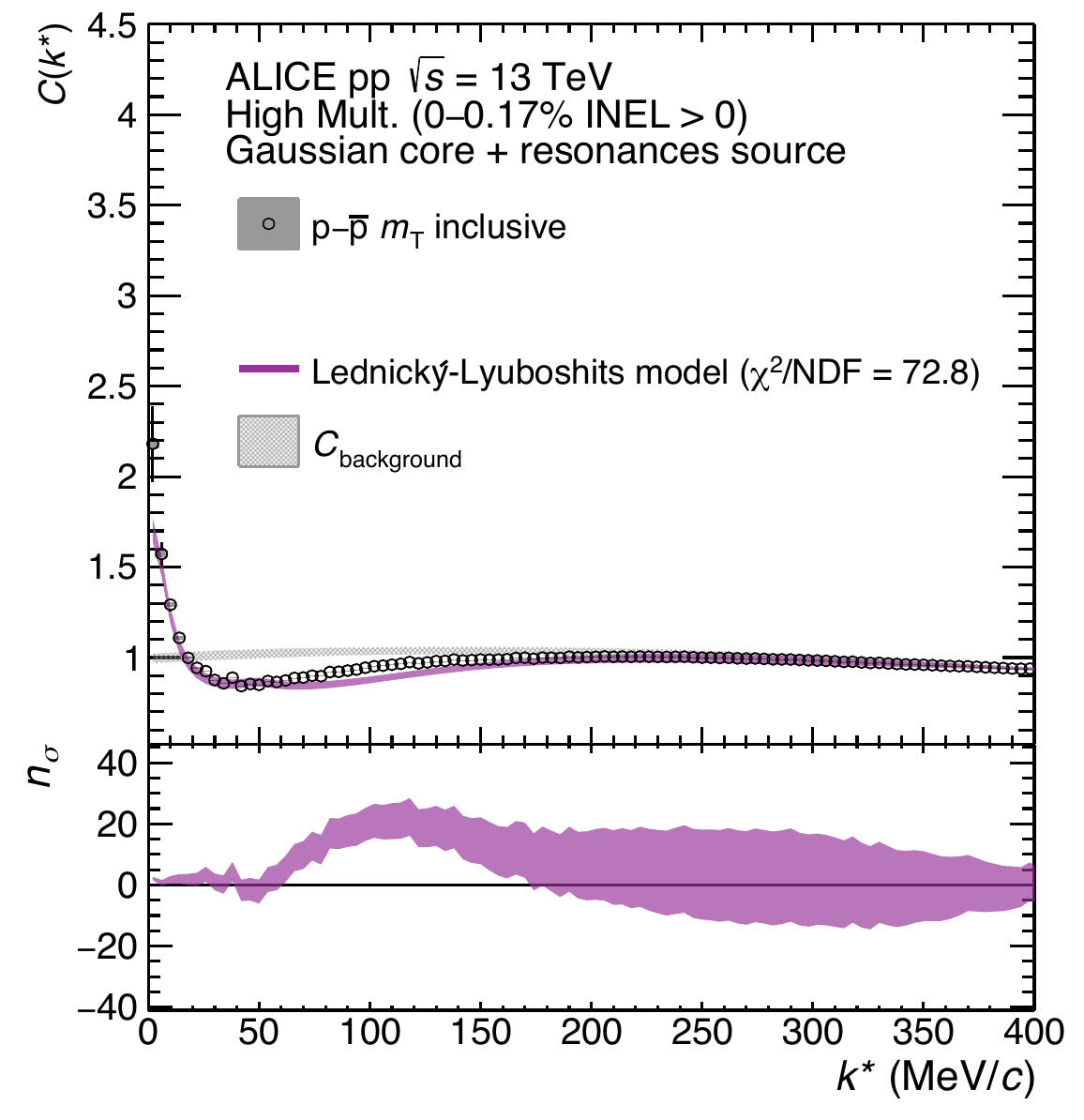}
    \caption{(Color online) Measured correlation function of \pap pairs. Statistical (bars) and systematic (boxes) uncertainties are shown separately. The results assuming the \Ledn model with Coulomb included, as in~\cite{WUTPaper}, are shown by the violet band. The scattering parameters used as input for the \Ledn calculations include only the \nan contribution as coupled-channel~\cite{PaPLednModel1,PaPLednModel2,PaPLednModel3}. The model completely underestimates the \ks region from 50 to 150 \MeVc. As can be seen in Fig.~\ref{fig:pApCFs}, the annihilation channels play a role in this intermediate region and a better description of the data is achieved when using the Migdal-Watson approximation to include them. This is a clear indication that the multi-meson channels are explicitly needed to model the current measured \pap correlation function. The \Ledn calculation also overestimates the coupling to the \nan channel, as can be seen from the large cusp structure at $\ks\approx50$ \MeVc not present in the data.  
    Lower panel: $n_\sigma$ deviation between data and model in terms of numbers of statistical standard deviations.}
    \label{fig:pAp_Lednicky}
\end{figure} 
%

\newpage

\section{The ALICE Collaboration}
\label{app:collab}
%
\begingroup
\small
\begin{flushleft}
S.~Acharya$^{\rm 143}$, 
D.~Adamov\'{a}$^{\rm 98}$, 
A.~Adler$^{\rm 76}$, 
J.~Adolfsson$^{\rm 83}$, 
G.~Aglieri Rinella$^{\rm 35}$, 
M.~Agnello$^{\rm 31}$, 
N.~Agrawal$^{\rm 55}$, 
Z.~Ahammed$^{\rm 143}$, 
S.~Ahmad$^{\rm 16}$, 
S.U.~Ahn$^{\rm 78}$
% ...

\section*{Affiliation notes}

$^{\rm I}$ Deceased\\
$^{\rm II}$ Also at: Italian National Agency for New Technologies, Energy and Sustainable Economic Development (ENEA), Bologna, Italy\\
$^{\rm III}$ Also at: Dipartimento DET del Politecnico di Torino, Turin, Italy\\
$^{\rm IV}$ Also at: M.V. Lomonosov Moscow State University, D.V. Skobeltsyn Institute of Nuclear, Physics, Moscow, Russia\\
$^{\rm V}$ Also at: Institute of Theoretical Physics, University of Wroclaw, Poland\\

\section*{Collaboration Institutes}

$^{1}$ A.I. Alikhanyan National Science Laboratory (Yerevan Physics Institute) Foundation, Yerevan, Armenia\\
$^{2}$ AGH University of Science and Technology, Cracow, Poland\\
$^{3}$ Bogolyubov Institute for Theoretical Physics, National Academy of Sciences of Ukraine, Kiev, Ukraine\\
$^{4}$ Bose Institute, Department of Physics  and Centre for Astroparticle Physics and Space Science (CAPSS), Kolkata, India\\
$^{5}$ Budker Institute for Nuclear Physics, Novosibirsk, Russia\\
$^{6}$ California Polytechnic State University, San Luis Obispo, California, United States\\
$^{7}$ Central China Normal University, Wuhan, China\\
$^{8}$ Centro de Aplicaciones Tecnol\'{o}gicas y Desarrollo Nuclear (CEADEN), Havana, Cuba\\
$^{9}$ Centro de Investigaci\'{o}n y de Estudios Avanzados (CINVESTAV), Mexico City and M\'{e}rida, Mexico\\
$^{10}$ Chicago State University, Chicago, Illinois, United States\\
$^{11}$ China Institute of Atomic Energy, Beijing, China\\
$^{12}$ Chungbuk National University, Cheongju, Republic of Korea\\
$^{13}$ Comenius University Bratislava, Faculty of Mathematics, Physics and Informatics, Bratislava, Slovakia\\
$^{14}$ COMSATS University Islamabad, Islamabad, Pakistan\\
$^{15}$ Creighton University, Omaha, Nebraska, United States\\
$^{16}$ Department of Physics, Aligarh Muslim University, Aligarh, India\\
$^{17}$ Department of Physics, Pusan National University, Pusan, Republic of Korea\\
$^{18}$ Department of Physics, Sejong University, Seoul, Republic of Korea\\
$^{19}$ Department of Physics, University of California, Berkeley, California, United States\\
$^{20}$ Department of Physics, University of Oslo, Oslo, Norway\\
$^{21}$ Department of Physics and Technology, University of Bergen, Bergen, Norway\\
$^{22}$ Dipartimento di Fisica dell'Universit\`{a} 'La Sapienza' and Sezione INFN, Rome, Italy\\
$^{23}$ Dipartimento di Fisica dell'Universit\`{a} and Sezione INFN, Cagliari, Italy\\
$^{24}$ Dipartimento di Fisica dell'Universit\`{a} and Sezione INFN, Trieste, Italy\\
$^{25}$ Dipartimento di Fisica dell'Universit\`{a} and Sezione INFN, Turin, Italy\\
$^{26}$ Dipartimento di Fisica e Astronomia dell'Universit\`{a} and Sezione INFN, Bologna, Italy\\
$^{27}$ Dipartimento di Fisica e Astronomia dell'Universit\`{a} and Sezione INFN, Catania, Italy\\
$^{28}$ Dipartimento di Fisica e Astronomia dell'Universit\`{a} and Sezione INFN, Padova, Italy\\
$^{29}$ Dipartimento di Fisica e Nucleare e Teorica, Universit\`{a} di Pavia, Pavia, Italy\\
$^{30}$ Dipartimento di Fisica `E.R.~Caianiello' dell'Universit\`{a} and Gruppo Collegato INFN, Salerno, Italy\\
$^{31}$ Dipartimento DISAT del Politecnico and Sezione INFN, Turin, Italy\\
$^{32}$ Dipartimento di Scienze e Innovazione Tecnologica dell'Universit\`{a} del Piemonte Orientale and INFN Sezione di Torino, Alessandria, Italy\\
$^{33}$ Dipartimento di Scienze MIFT, Universit\`{a} di Messina, Messina, Italy\\
$^{34}$ Dipartimento Interateneo di Fisica `M.~Merlin' and Sezione INFN, Bari, Italy\\
$^{35}$ European Organization for Nuclear Research (CERN), Geneva, Switzerland\\
$^{36}$ Faculty of Electrical Engineering, Mechanical Engineering and Naval Architecture, University of Split, Split, Croatia\\
$^{37}$ Faculty of Engineering and Science, Western Norway University of Applied Sciences, Bergen, Norway\\
$^{38}$ Faculty of Nuclear Sciences and Physical Engineering, Czech Technical University in Prague, Prague, Czech Republic\\
$^{39}$ Faculty of Science, P.J.~\v{S}af\'{a}rik University, Ko\v{s}ice, Slovakia\\
$^{40}$ Frankfurt Institute for Advanced Studies, Johann Wolfgang Goethe-Universit\"{a}t Frankfurt, Frankfurt, Germany\\
$^{41}$ Fudan University, Shanghai, China\\
$^{42}$ Gangneung-Wonju National University, Gangneung, Republic of Korea\\
$^{43}$ Gauhati University, Department of Physics, Guwahati, India\\
$^{44}$ Helmholtz-Institut f\"{u}r Strahlen- und Kernphysik, Rheinische Friedrich-Wilhelms-Universit\"{a}t Bonn, Bonn, Germany\\
$^{45}$ Helsinki Institute of Physics (HIP), Helsinki, Finland\\
$^{46}$ High Energy Physics Group,  Universidad Aut\'{o}noma de Puebla, Puebla, Mexico\\
$^{47}$ Hiroshima University, Hiroshima, Japan\\
$^{48}$ Hochschule Worms, Zentrum  f\"{u}r Technologietransfer und Telekommunikation (ZTT), Worms, Germany\\
$^{49}$ Horia Hulubei National Institute of Physics and Nuclear Engineering, Bucharest, Romania\\
$^{50}$ Indian Institute of Technology Bombay (IIT), Mumbai, India\\
$^{51}$ Indian Institute of Technology Indore, Indore, India\\
$^{52}$ Indonesian Institute of Sciences, Jakarta, Indonesia\\
$^{53}$ INFN, Laboratori Nazionali di Frascati, Frascati, Italy\\
$^{54}$ INFN, Sezione di Bari, Bari, Italy\\
$^{55}$ INFN, Sezione di Bologna, Bologna, Italy\\
$^{56}$ INFN, Sezione di Cagliari, Cagliari, Italy\\
$^{57}$ INFN, Sezione di Catania, Catania, Italy\\
$^{58}$ INFN, Sezione di Padova, Padova, Italy\\
$^{59}$ INFN, Sezione di Pavia, Pavia, Italy\\
$^{60}$ INFN, Sezione di Roma, Rome, Italy\\
$^{61}$ INFN, Sezione di Torino, Turin, Italy\\
$^{62}$ INFN, Sezione di Trieste, Trieste, Italy\\
$^{63}$ Inha University, Incheon, Republic of Korea\\
$^{64}$ Institute for Gravitational and Subatomic Physics (GRASP), Utrecht University/Nikhef, Utrecht, Netherlands\\
$^{65}$ Institute for Nuclear Research, Academy of Sciences, Moscow, Russia\\
$^{66}$ Institute of Experimental Physics, Slovak Academy of Sciences, Ko\v{s}ice, Slovakia\\
$^{67}$ Institute of Physics, Homi Bhabha National Institute, Bhubaneswar, India\\
$^{68}$ Institute of Physics of the Czech Academy of Sciences, Prague, Czech Republic\\
$^{69}$ Institute of Space Science (ISS), Bucharest, Romania\\
$^{70}$ Institut f\"{u}r Kernphysik, Johann Wolfgang Goethe-Universit\"{a}t Frankfurt, Frankfurt, Germany\\
$^{71}$ Instituto de Ciencias Nucleares, Universidad Nacional Aut\'{o}noma de M\'{e}xico, Mexico City, Mexico\\
$^{72}$ Instituto de F\'{i}sica, Universidade Federal do Rio Grande do Sul (UFRGS), Porto Alegre, Brazil\\
$^{73}$ Instituto de F\'{\i}sica, Universidad Nacional Aut\'{o}noma de M\'{e}xico, Mexico City, Mexico\\
$^{74}$ iThemba LABS, National Research Foundation, Somerset West, South Africa\\
$^{75}$ Jeonbuk National University, Jeonju, Republic of Korea\\
$^{76}$ Johann-Wolfgang-Goethe Universit\"{a}t Frankfurt Institut f\"{u}r Informatik, Fachbereich Informatik und Mathematik, Frankfurt, Germany\\
$^{77}$ Joint Institute for Nuclear Research (JINR), Dubna, Russia\\
$^{78}$ Korea Institute of Science and Technology Information, Daejeon, Republic of Korea\\
$^{79}$ KTO Karatay University, Konya, Turkey\\
$^{80}$ Laboratoire de Physique des 2 Infinis, Ir\`{e}ne Joliot-Curie, Orsay, France\\
$^{81}$ Laboratoire de Physique Subatomique et de Cosmologie, Universit\'{e} Grenoble-Alpes, CNRS-IN2P3, Grenoble, France\\
$^{82}$ Lawrence Berkeley National Laboratory, Berkeley, California, United States\\
$^{83}$ Lund University Department of Physics, Division of Particle Physics, Lund, Sweden\\
$^{84}$ Moscow Institute for Physics and Technology, Moscow, Russia\\
$^{85}$ Nagasaki Institute of Applied Science, Nagasaki, Japan\\
$^{86}$ Nara Women{'}s University (NWU), Nara, Japan\\
$^{87}$ National and Kapodistrian University of Athens, School of Science, Department of Physics , Athens, Greece\\
$^{88}$ National Centre for Nuclear Research, Warsaw, Poland\\
$^{89}$ National Institute of Science Education and Research, Homi Bhabha National Institute, Jatni, India\\
$^{90}$ National Nuclear Research Center, Baku, Azerbaijan\\
$^{91}$ National Research Centre Kurchatov Institute, Moscow, Russia\\
$^{92}$ Niels Bohr Institute, University of Copenhagen, Copenhagen, Denmark\\
$^{93}$ Nikhef, National institute for subatomic physics, Amsterdam, Netherlands\\
$^{94}$ NRC Kurchatov Institute IHEP, Protvino, Russia\\
$^{95}$ NRC \guillemotleft Kurchatov\guillemotright  Institute - ITEP, Moscow, Russia\\
$^{96}$ NRNU Moscow Engineering Physics Institute, Moscow, Russia\\
$^{97}$ Nuclear Physics Group, STFC Daresbury Laboratory, Daresbury, United Kingdom\\
$^{98}$ Nuclear Physics Institute of the Czech Academy of Sciences, \v{R}e\v{z} u Prahy, Czech Republic\\
$^{99}$ Oak Ridge National Laboratory, Oak Ridge, Tennessee, United States\\
$^{100}$ Ohio State University, Columbus, Ohio, United States\\
$^{101}$ Petersburg Nuclear Physics Institute, Gatchina, Russia\\
$^{102}$ Physics department, Faculty of science, University of Zagreb, Zagreb, Croatia\\
$^{103}$ Physics Department, Panjab University, Chandigarh, India\\
$^{104}$ Physics Department, University of Jammu, Jammu, India\\
$^{105}$ Physics Department, University of Rajasthan, Jaipur, India\\
$^{106}$ Physikalisches Institut, Eberhard-Karls-Universit\"{a}t T\"{u}bingen, T\"{u}bingen, Germany\\
$^{107}$ Physikalisches Institut, Ruprecht-Karls-Universit\"{a}t Heidelberg, Heidelberg, Germany\\
$^{108}$ Physik Department, Technische Universit\"{a}t M\"{u}nchen, Munich, Germany\\
$^{109}$ Politecnico di Bari and Sezione INFN, Bari, Italy\\
$^{110}$ Research Division and ExtreMe Matter Institute EMMI, GSI Helmholtzzentrum f\"ur Schwerionenforschung GmbH, Darmstadt, Germany\\
$^{111}$ Russian Federal Nuclear Center (VNIIEF), Sarov, Russia\\
$^{112}$ Saha Institute of Nuclear Physics, Homi Bhabha National Institute, Kolkata, India\\
$^{113}$ School of Physics and Astronomy, University of Birmingham, Birmingham, United Kingdom\\
$^{114}$ Secci\'{o}n F\'{\i}sica, Departamento de Ciencias, Pontificia Universidad Cat\'{o}lica del Per\'{u}, Lima, Peru\\
$^{115}$ St. Petersburg State University, St. Petersburg, Russia\\
$^{116}$ Stefan Meyer Institut f\"{u}r Subatomare Physik (SMI), Vienna, Austria\\
$^{117}$ SUBATECH, IMT Atlantique, Universit\'{e} de Nantes, CNRS-IN2P3, Nantes, France\\
$^{118}$ Suranaree University of Technology, Nakhon Ratchasima, Thailand\\
$^{119}$ Technical University of Ko\v{s}ice, Ko\v{s}ice, Slovakia\\
$^{120}$ The Henryk Niewodniczanski Institute of Nuclear Physics, Polish Academy of Sciences, Cracow, Poland\\
$^{121}$ The University of Texas at Austin, Austin, Texas, United States\\
$^{122}$ Universidad Aut\'{o}noma de Sinaloa, Culiac\'{a}n, Mexico\\
$^{123}$ Universidade de S\~{a}o Paulo (USP), S\~{a}o Paulo, Brazil\\
$^{124}$ Universidade Estadual de Campinas (UNICAMP), Campinas, Brazil\\
$^{125}$ Universidade Federal do ABC, Santo Andre, Brazil\\
$^{126}$ University of Cape Town, Cape Town, South Africa\\
$^{127}$ University of Houston, Houston, Texas, United States\\
$^{128}$ University of Jyv\"{a}skyl\"{a}, Jyv\"{a}skyl\"{a}, Finland\\
$^{129}$ University of Kansas, Lawrence, Kansas, United States\\
$^{130}$ University of Liverpool, Liverpool, United Kingdom\\
$^{131}$ University of Science and Technology of China, Hefei, China\\
$^{132}$ University of South-Eastern Norway, Tonsberg, Norway\\
$^{133}$ University of Tennessee, Knoxville, Tennessee, United States\\
$^{134}$ University of the Witwatersrand, Johannesburg, South Africa\\
$^{135}$ University of Tokyo, Tokyo, Japan\\
$^{136}$ University of Tsukuba, Tsukuba, Japan\\
$^{137}$ Universit\'{e} Clermont Auvergne, CNRS/IN2P3, LPC, Clermont-Ferrand, France\\
$^{138}$ Universit\'{e} de Lyon, CNRS/IN2P3, Institut de Physique des 2 Infinis de Lyon , Lyon, France\\
$^{139}$ Universit\'{e} de Strasbourg, CNRS, IPHC UMR 7178, F-67000 Strasbourg, France, Strasbourg, France\\
$^{140}$ Universit\'{e} Paris-Saclay Centre d'Etudes de Saclay (CEA), IRFU, D\'{e}partment de Physique Nucl\'{e}aire (DPhN), Saclay, France\\
$^{141}$ Universit\`{a} degli Studi di Foggia, Foggia, Italy\\
$^{142}$ Universit\`{a} di Brescia, Brescia, Italy\\
$^{143}$ Variable Energy Cyclotron Centre, Homi Bhabha National Institute, Kolkata, India\\
$^{144}$ Warsaw University of Technology, Warsaw, Poland\\
$^{145}$ Wayne State University, Detroit, Michigan, United States\\
$^{146}$ Westf\"{a}lische Wilhelms-Universit\"{a}t M\"{u}nster, Institut f\"{u}r Kernphysik, M\"{u}nster, Germany\\
$^{147}$ Wigner Research Centre for Physics, Budapest, Hungary\\
$^{148}$ Yale University, New Haven, Connecticut, United States\\
$^{149}$ Yonsei University, Seoul, Republic of Korea\\

\bigskip 

\end{flushleft} 
\endgroup
  
\end{document}